\documentclass{elsart}

\usepackage{graphics}

\usepackage{graphicx}

\usepackage{amssymb}

\usepackage[dvips]{epsfig}                   

 \usepackage{lineno}

\newcommand*{\no}{\noindent}
\newcommand*{\bea}{\begin{eqnarray}}
\newcommand*{\eea}{\end{eqnarray}}

\newcommand*{\pref}[1]{(\ref{#1})}

\newcommand*{\mn}{{\mu\nu}}

\newcommand*{\nn}{\nonumber}

\newcommand{\cP}{\mathcal{P}}

 \newcommand{\zr}[1]{\mbox{\hspace*{#1em}}}

 \newcommand{\ZZ}{\mbox{\sf Z\zr{-0.45}Z}}

\def\di{\displaystyle}

\def\bg{\begin{eqnarray}\begin{array}{rcl}\displaystyle}
\def\eg{\end{array} &\di    &\di   \end{eqnarray}}
\def\bm#1{\begin{eqnarray}\begin{array}{#1}\di}
\def\bmo#1{\begin{eqnarray*}\begin{array}{#1}\di}
\def\bml#1#2{\begin{eqnarray}\begin{array}{#1}\label{#2}\di}
\def\bgo{\begin{eqnarray*}\begin{array}{rcl}\displaystyle}
\def\ego{\end{array} &\di    &\di \nonumber  \end{eqnarray*}}
\def\no{\nonumber}
\def\btensor#1#2{\renew\left#1\begin{array}{#2}\di}
\def\brtensor#1#2#3{\ren#3\left#1\begin{array}{#2}}
\def\botensor#1#2{\renew\left#1\begin{array}{#2}}
\def\etensor#1{\end{array}\right#1}

\def\eq#1{(\ref{#1})}
\def\Eq#1{Eq.~(\ref{#1})}

\def\s0#1#2{\mbox{\small{$ \frac{#1}{#2} $}}}
\def\0#1#2{\frac{#1}{#2}}

\def\CO{{\mathcal O}}
\def\CP{{\mathcal P}}

\renewcommand{\theequation}{\arabic{section}.\arabic{equation}}

\def\ren#1{\renewcommand{\arraystretch}{#1}}

\def\renew{\renewcommand{\arraystretch}{1}}
\renewcommand{\jot}{10pt}

\begin{document}

\begin{frontmatter}

\title{Large volume behaviour of Yang-Mills propagators}


\author[a]{Christian S.~Fischer,}
\author[b,c]{Axel Maas,}
\author[d]{Jan M.~Pawlowski,}
\author[e]{and Lorenz von Smekal,}

\address[a]{Institut f\"ur Kernphysik, University of Technology,
  Schlossgartenstra{\ss}e 9,\\ D-64289 Darmstadt, Germany}

\address[b]{Instituto de F\'isica de S\~ao Carlos, University of
  S\~ao Paulo, C.\ P.\ 369,\\ 13560-970 S\~ao Carlos, SP, Brazil}

\address[c]{Gesellschaft f\"ur Schwerionenforschung mbH,
  Planckstra{\ss}e 1,\\
  D-64291 Darmstadt, Germany}

\address[d]{Institut f\"ur Theoretische Physik, University of
  Heidelberg, Philosophenweg 16, D-62910 Heidelberg, Germany}

\address[e]{Centre for the Subatomic Structure of Matter, School of
  Chemistry and Physics, The University of Adelaide, SA 5005,
  Australia}



\begin{abstract}
  We investigate finite volume effects in the propagators of Landau
  gauge Yang-Mills theory using Dyson-Schwinger equations on a
  4-dimensional torus.  In particular, we demonstrate explicitly how
  the solutions for the gluon and the ghost propagator tend towards
  their respective infinite volume forms in the corresponding limit.
  This solves an important open problem of previous studies where the
  infinite volume limit led to an apparent mismatch, especially of the
  infrared behaviour, between torus extrapolations and the existing
  infinite volume solutions obtained in 4-dimensional Euclidean
  space-time.  However, the correct infinite volume limit is
  approached rather slowly. The typical scales necessary to see the
  onset of the leading infrared behaviour emerging already imply
  volumes of at least 10 to 15 fm in lengths. To reliably extract the
  infrared exponents of the infinite volume solutions requires even
  much larger ones.  While the volumes in the Monte-Carlo simulations
  available at present are far too small to facilitate that, we obtain
  a good qualitative agreement of our torus solutions with recent
  lattice data in comparable volumes.

\end{abstract}

\end{frontmatter}

\newpage

\section{Introduction}

The infrared behaviour of the QCD Green's functions is known to
contain essential information about the realisation of confinement in
the covariant continuum formulation of QCD in terms of local field
systems \cite{Alkofer:2000wg}. To extract this information from
studies of correlation functions in a finite volume it is most
important to address the volume dependence of the long-range, {\it
  i.e.} infrared, behaviour of these correlations. It is only when
this dependence is under control that firm conclusions can be drawn
from infinite volume extrapolations.

In this paper we revisit the elementary 2-point correlation functions,
{\it i.e.} the propagators, of pure Yang-Mills theory without quarks
in a finite volume. Using covariant gauges, and here in particular the
Landau gauge, these are the gluon and the ghost propagators. These are
perhaps the most important examples of how Green's functions relate to
confinement in this formulation. Their infrared behaviour is
sensitive to confinement according to the scenarios of Kugo and Ojima
\cite{Kugo1979,Nakanishi1990}, and of Gribov and Zwanziger
\cite{Gribov:1977wm,Zwanziger:1993dh}. In Landau gauge these scenarios 
predict an infrared vanishing gluon propagator and an infrared
enhanced ghost propagator (the latter was pointed out in
\cite{Kugo1995}; for a review, see \cite{Alkofer:2000wg}).

Qualitatively, finite-volume effects are expected to distort the
infrared properties of 2-point correlation functions for momenta 
approaching $2\pi/L$, where $L$ is the finite length of the system in
the corresponding direction. For pure Yang-Mills theory the momentum
scale for confinement is set by $\Lambda_{\rm QCD}$, which is of the order of 
200 -- 400 MeV, somewhat dependent on the scheme.\footnote{{\it E.g.}, 
  $\Lambda_{\overline{MS}}\approx 240 $ MeV for $N_f = 0$ from the
  lattice determination by the ALPHA Collaboration \cite{Sommer2006}.}
This implies that for lengths around $L_c=2\pi/\Lambda_{\rm 
  QCD}$, of the order $3$ -- $6$ fm, the finite volume will already have a 
considerable effect on the infrared behaviour of these
confinement-sensitive correlations. Eventually, of course, when
further decreasing the volume down to the sub-Fermi regime,  
confinement will be lost altogether. 

In present Monte-Carlo simulations, using lattice Landau gauge, the
gluon and ghost propagators of pure $SU(2)$ and $SU(3)$ gauge theory
in four dimensions are obtained in volumes of the order of $L_c$ 
in length, in some cases even somewhat larger than $L_c$  
\cite{Bonnet2000,Bonnet2001,Cucchieri:2006xi,Sternbeck:2006cg}.
However, despite clear indications in favour of the qualitative
infrared behaviour of both the gluon and the ghost propagator, as
predicted by continuum studies, some quantitative discrepancies still
remain. In particular, apart from some indications from the volume
scaling of the zero momentum gluon propagator, a systematic and
statistically significant verification of its infrared suppression
from lattice data in 4 dimensions is still lacking.\footnote{This is
  different in Coulomb gauge  and interpolating gauges, where a
  stronger suppression is expected from the Gribov-Zwanziger scenario,
  and where this infrared suppression of the gluon propagator {\em is}
  observed on the lattice already in rather small volumes
  \cite{CucchieriCoulomb,int}.}

As far as the continuum studies are concerned, on the other hand, we
are nowadays in the quite comfortable situation that a variety of
different non-perturbative approaches all lead to the same infrared
behaviour for the propagators in the infinite volume limit. These
include studies of their Dyson-Schwinger Equations (DSEs)
\cite{Lerche:2002ep}, of the Fokker-Planck type diffusion equations of
Stochastic Quantisation \cite{Zwanziger:2001kw}, and of the Functional
Renormalisation Group Equations (FRGEs)~\cite{Pawlowski:2003hq}.

In order to understand the origin of the
remaining discrepancies observed between the functional methods in the
continuum and the lattice Landau gauge simulations, it is an obvious
and necessary step forward to adapt the continuum methods to finite
volumes. The techniques to solve Dyson-Schwinger equations on a
finite four dimensional torus with periodic boundary conditions have
recently been developed and applied to the various propagator DSEs of
Landau gauge QCD \cite{Fischer,Fischer:2005ui,Fischer:2005nf}.
The infinite volume limit, however, remained unclear in these studies. 
Here we specifically address this question and explicitly demonstrate
how the infrared behaviour known from the previous studies of 
functional methods is approached in the infinite volume limit also 
by the DSE solutions on a finite torus.

In fact, this is to be expected. The perturbative and the confining
regime are separated by a cross-over, the scale of which is
$\Lambda_{\rm QCD}$, the dynamical scale. As pointed out in previous
infrared studies
\cite{Lerche:2002ep,Pawlowski:2003hq,Pawlowski:2005xe}, for the investigation
of the infrared regime of the correlation functions we have to
consider $\Lambda_{\rm QCD}$ an ultraviolet scale which is large
compared to all external momenta $p$ involved, {\it i.e.}, $ p \ll
\Lambda_{\rm QCD}$. As mentioned above, for the emergence of continuum
infrared physics we need in addition that $2\pi/L \ll p$.  To reliably
extract infrared critical exponents from finite volume studies we
therefore need a clear separation of scales with a sufficient number
of different momentum values all in the range
\begin{equation}
      \0{2\pi}{L}\ll p\ll \Lambda_{\rm QCD}\; . \label{eq:valrange}
\end{equation}
Indeed, the same scale separation was used in the infrared analysis
of the functional RGEs \cite{Pawlowski:2003hq,Pawlowski:2005xe}, 
where an explicit infrared momentum cut-off $k$ is introduced and plays a
role analogous to that by the finite volume. In the functional 
RG infrared-studies it is precisely this regime,  $k \ll p \ll
\Lambda_{\rm QCD}$, that is required to lead to the same infrared
behaviour as in the other functional continuum investigations without
such infrared cut-off \cite{Lerche:2002ep,Zwanziger:2001kw}. In the
available studies of the DSEs on a symmetric hypertorus, on the other
hand, this same limit has not been explicitly obtained as yet.
In the present work we fill this gap by presenting a refined infrared
analysis.

Because of the necessary separation of scales (\ref{eq:valrange}), the
actual lattice sizes $L$ necessary to achieve this separation for a
sufficiently wide range of momentum values need to be much larger than
$L_c$. From our numerical analysis we estimate that this might well
require lengths $L$ of the order of 40 fm, so that reliable lattice
determinations of the infrared exponents of the gluon and ghost
propagators on reasonably fine lattices are likely to remain
unfeasible for some time to come.  Corresponding $SU(2)$ simulations
on rather large lattices in three dimensions
\cite{Cucchieri:2003di} confirm this by showing a slow but
steady convergence towards the three dimensional infinite
volume limit, as predicted by DSE studies
\cite{Zwanziger:2001kw,Maas:2004se}.

Meanwhile, we explicitly demonstrate this convergence here for the DSE
solutions in a finite 4-dimensional volume by investigating the
behaviour of the propagators on length scales in the range given by
(\ref{eq:valrange}). Sec.~\ref{sec:sana} is devoted to analytic
results.  After a brief summary of the infrared behaviour of the
propagators from infinite volume studies in Sec.~\ref{sec:infvol}, we
discuss the torus DSEs and their proper renormalisation in
Sec.~\ref{sec:torusDSEs}. The question of zero modes is addressed in
Sec.~\ref{sec:sszero}. The infrared behaviour of the torus-DSE
solutions and the volume dependence of the low-momentum modes are
studied in Sec.~\ref{sec:ssvols}.  Our numerical results are presented
in Sec.~\ref{sec:snum}. This contains a brief description of the
general procedure and the implementation of the renormalisation
procedure in Secs.~\ref{sec:mom} and \ref{sec:renorm}, and a detailed
discussion of the numerical solutions for the gluon and ghost
propagators in various volumes in Sec.~\ref{sec:props}. In
Sec.~\ref{sec:asym} we present fits and extrapolations to model the
approach towards the infinite volume limit, and in
Sec.~\ref{sec:lattice} we compare our solutions to the results of
lattice simulations with similar volumes.

Our conclusions are provided in Sec.~\ref{sec:sconc}. 
We summarise that the presently available lattice data is consistent
with the continuum results, if the consequences of finite-volume
effects are properly taken into account. Moreover, the infinite-volume
extrapolations of our torus-DSE solutions indicate that volumes of
lengths of the order of 40 fm might be necessary to explicitly verify the
infrared behaviour of the gluon and ghost propagators in the various   
functional continuum approaches, and to obtain reliable estimates from
lattice simulations of the critical exponents indicative of this
conformal infrared behaviour of the pure gauge theory in the infinite
volume, the 4-dimensional Euclidean space-time.

\section{Analytic Results}\label{sec:sana}

\subsection{Dyson-Schwinger equations at infinite volume}
\label{sec:infvol}

The Dyson-Schwinger equations of QCD in 4-dimensional Euclidean space-time 
have been used very successfully to determine the Green's functions in
particular of the Landau gauge \cite{Alkofer:2000wg,Fischer:rev}.

For the propagators of the pure gauge theory in the Landau gauge the
full momentum dependence is parametrised by two dimensionless dressing
functions $Z$ and $G$ as follows,
 \begin{eqnarray}
   D^{ab}_\mn(p)&=&\delta^{ab}\left(\delta_\mn-
     \frac{p_\mu p_\nu}{p^2}
   \right)
   \frac{Z(p)}{p^2} \; ,\nn\\
   D^{ab}(p^2)&=&-\delta^{ab}\frac{G(p)}{p^2} \; , \nn  
 \end{eqnarray} 
 \no 
 where $D_\mn^{ab}$ is the gluon propagator and $D^{ab}$ is the
 propagator of the Faddeev-Popov ghosts.

 The infrared behaviour of these dressing functions $Z$ and $G$ for
 gluons and ghosts is determined by one unique infrared exponent $0<
 \kappa < 1$ \cite{vonSmekal1997,vonSmekal1998},
\begin{eqnarray} 
  Z_\mathrm{IR}(p^2) \sim (p^2)^{2\kappa}\,, \ \ \ \
  G_\mathrm{IR}(p^2) \sim (p^2)^{-\kappa} \, .\label{powers} 
\end{eqnarray} 
The exact value of this exponent $\kappa $ depends only on the
infrared behaviour of a single invariant function $A(k^2;p^2,q^2)$
\cite{Lerche:2002ep} which multiplies the tree-level structure in the
full ghost-gluon vertex. Here $k^2 $ denotes the gluon momentum and
$p^2$ and $q^2$ are the (anti)ghost momenta. In ghost/anti-ghost
symmetric gauges, such as the Landau gauge, $A$ is symmetric under the
exchange of the ghost momenta, $A(k^2;q^2,p^2) = A(k^2;p^2,q^2)$.
If this function is furthermore assumed to be finite and regular in the
origin at $k^2 = p^2 = q^2 = 0$, then the value of the infrared
exponent $\kappa $ is \cite{Lerche:2002ep}
\begin{eqnarray} 
  \kappa =\kappa_c = \frac{93-\sqrt{1201}}{98} \approx 0.595 
  \label{kappa} \; .
  \label{kappa_c}
\end{eqnarray} 
This regularity assumption is all that is needed to obtain this
otherwise exact result. The same value of $\kappa $ was obtained
independently at the same time using a bare vertex (which is trivially
regular) from the time-independent (equilibrium version of the)
diffusion equation of Stochastic Quantisation \cite{Zwanziger:2001kw}.
In the corresponding truncation of the functional RGEs a range of values
roughly between $0.54 $ and $\kappa_c$ is possible with
(\ref{kappa_c}) representing the special value obtained for an
optimised flow \cite{Pawlowski:2003hq,Pawlowski:2005xe}.

The same regularity assumption implicitly underlies all these studies
and produces the unique result (\ref{kappa_c}). Other values for
$\kappa $ in the range $1/2 \le \kappa < 1 $ are possible, if the
tree-level structure $A$ of the ghost-gluon vertex has infrared
divergences associated with any of its legs \cite{Lerche:2002ep}.

So how justified is this regularity assumption? It is relatively easy
to see that $ A(p^2;0,p^2) = A(p^2;p^2,0) = 1$ in Landau gauge. An
independent and different argument, from the non-renormalisation of
the ghost-gluon vertex in Landau gauge \cite{Taylor:1971ff},
perturbatively at all orders, implies $A(p^2; p^2, p^2) = 1 $ in a
symmetric momentum subtraction scheme. If this argument remains true
beyond perturbation theory it fixes the infrared limit of $A$ along a
second direction. Lattice simulations of the ghost-gluon vertex have
so far tested the line $A(0;p^2,p^2)$ in $SU(2)$ and $SU(3)$ for
momenta $p$ down to values of the order of $\Lambda_{\rm QCD}$, and
obtained results consistent with $A(0;p^2,p^2) = 1$ with no systematic
indications of any momentum dependence along this third direction
either \cite{Cucchieri:2004sq,Sternbeck:2005qj,%
 Sternbeck:2006rd}.\footnote{For more results in $SU(2)$, including
  some different kinematic regimes, see \cite{Cucchieri:2006tf}.}   
While none of this can prove the infrared regularity of this
ghost-gluon vertex structure, there is certainly no evidence to the
contrary at present.

Probably more importantly, however, there is a self-consistent scheme
to solve the full hierarchy of DSEs in the pure gauge theory
asymptotically in the infrared \cite{afl}. Remember that the complete
system of DSEs forms an infinite set of coupled non-linear integral
equations between all $n$-point Green's functions. The self-consistent
infrared-asymptotic solution to the whole tower of $n$-point Green's
functions obtained in \cite{afl} reflects the infrared fixed-point
behaviour of the pure gauge theory. The ghost-gluon vertex {\em is}
regular in the infrared in this solution, which incorporates
(\ref{powers}) and leads to a simple counting scheme for generalised
power laws of this kind for all $n$-point Green's functions. Moreover,
it has recently been proven that this solution is unique among this
class of parametrisations of a conformal infrared behaviour
\cite{Fischer:2006vf}.

Because of the non-renormalisation of the ghost-gluon vertex in Landau
gauge, the ghost and gluon dressing functions provide a
non-perturbative definition of the running coupling in this gauge via
the RG invariant product \cite{vonSmekal1998}
\begin{equation}
  \alpha(p^2) = \alpha(\mu^2) \, G(p^2)^2 \, Z(p^2) \,, \label{coupling} 
\end{equation}
where $\mu^2$ is the renormalisation point. It approaches a finite
positive value for $p^2 \to 0$ in four dimensions for all $0 < \kappa
<1 $. The value of this infrared fixed-point depends on $\kappa $,
{\it i.e.}, on the infrared properties of the ghost-gluon vertex (it
vanishes at both end points of this interval, for $\kappa \to 0$ and
for $\kappa \to 1$).  Moreover, the maximal value of the coupling at
the infrared fixed-point is that obtained for the infrared-regular
ghost-gluon vertex, with $\kappa = \kappa_c$ from (\ref{kappa}), which
yields \cite{Lerche:2002ep},
\begin{eqnarray} 
  \alpha(0) = \alpha_c = \frac{8 \pi}{N_c}\; 
  \frac{\Gamma^2(\kappa_c-1) \, \Gamma(4-2\kappa_c) }
  {\Gamma^2(-\kappa_c) \, \Gamma(2\kappa_c-1)} \,\approx\, 
    \frac{4 \pi}{N_c} \; 0.709 \, \approx \, 8.9/N_c
  \label{IRfixed}  \; ,
\end{eqnarray}
for $N_c$ colours. This implies $\alpha_c \approx 4.46$ for
$SU(2)$ and $\alpha_c  \approx 2.97 $ for $SU(3)$.

\begin{figure}[t]
\centerline{\includegraphics[width=12cm]{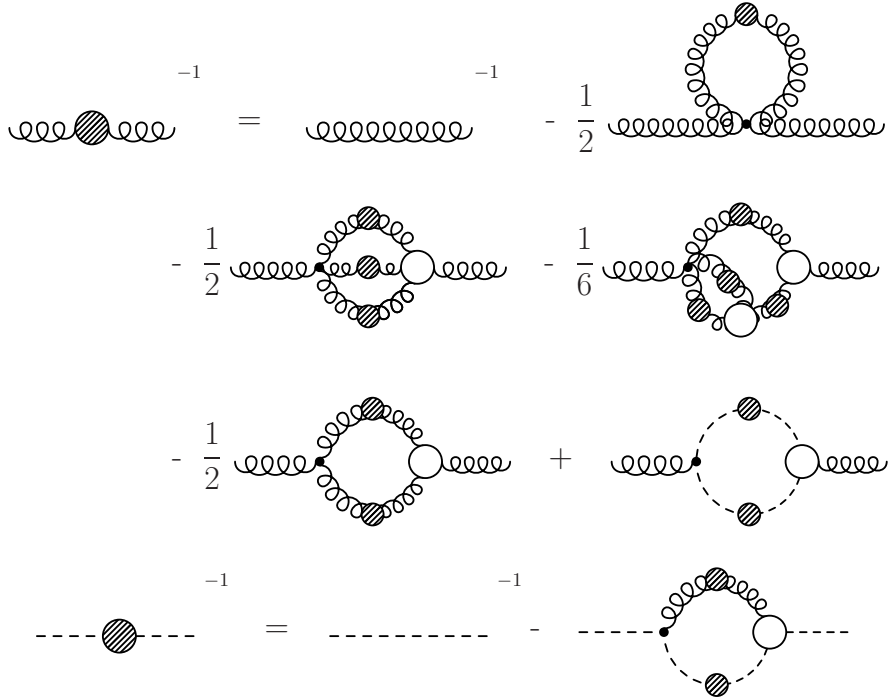}}
\caption{\label{sys}The propagator DSEs. Curly lines are gluons and
  dotted lines are ghosts. Lines with a large filled dot represent
  full propagators.  Vertices with a small dot are bare and with a
  large dot are full and constructed in this truncation.}
\end{figure} 

Numerical solutions of the DSEs can only be obtained in specific
truncation schemes. For the coupled system of ghost and gluon
propagator DSEs, given diagrammatically in Fig.~\ref{sys}, one such
scheme has been defined in \cite{Fischer,fa}. It uses a bare
ghost-gluon vertex, a choice well justified by both lattice studies
\cite{Cucchieri:2004sq,Sternbeck:2005qj,Sternbeck:2006rd,%
  Cucchieri:2006tf} as well as calculations using the vertex DSE
\cite{Schleifenbaum:2004id}.  Furthermore an Ansatz for the
three-gluon vertex has been used that ensures the correct infrared and
ultraviolet limits of the solutions for the ghost and gluon
propagators. Contributions involving the four-gluon interaction have
been neglected, a choice which only affects the intermediate momentum
regime \cite{Alkofer:2000wg,Fischer:rev}.  The resulting numerical
solutions of the ghost-gluon system have been reported in \cite{fa}
and are included here again in Sec.~\ref{sec:snum}.  In the following
we will denote these solutions as 'infinite volume solutions' as
opposed to the 'torus solutions' discussed in the next sections. Note
that for the latter we assume the continuum limit in a finite volume
even though, strictly speaking, our numerical solutions in
Sec.~\ref{sec:snum} are obtained on a discrete and finite set of
momentum values. The continuum limit thereby corresponds to removing
the ultraviolet cut-off, of course. In Sec.~\ref{sec:props} we will
verify that the residual cut-off dependences of the propagators are
negligible.  Unlike lattice simulations, where disentangling
discretisation errors and finite volume effects can be a rather
challenging task, it is therefore not necessary to study the continuum
limit more carefully for the DSE solutions, provided a consistent
renormalisation scheme is implemented.

\subsection{Dyson-Schwinger equations on a torus}
\label{sec:torusDSEs}

For the formulation of the Dyson-Schwinger equations of Fig.~\ref{sys}
in a 4-dimensional hypercubic volume $L^4$ of length $L$ in all
directions, with periodic boundary conditions, the momentum integrals
in the infinite volume DSEs are replaced by sums,
\begin{equation}
  \int \frac{d^4q}{(2\pi)^4}\, (\dots)  \longrightarrow 
  \frac{1}{L^4} \sum_{\mbox{\footnotesize $n \in  \ZZ^4 $}}\, (\dots) 
  \; ,\label{eq:torusints}
\end{equation}
where the four dimensional vector of integers  $n\in \ZZ^4$ labels the
discrete momentum values $q_n = (2\pi/L) \, n$. In the
following we truncate the DSEs shown in Fig.~\ref{sys}
such that the diagrams with (bare) four-gluon vertices are neglected,
analogous to the truncation scheme used in the infinite volume
computations as mentioned above. Denoting the gluon and ghost dressing
functions by $Z_L$ and $G_L$ in the finite volume, their corresponding
DSEs become, 
\begin{eqnarray} \label{eq:gluon} \frac{1}{Z_L(k)} & =& Z_3 -
  \frac{g^2 \, N_c \, \tilde Z_1}{3} \0{1}{L^4}\sum_{n} \frac{q_n
    \cP(k) p_n\,\, G_L(q_n)\, G_L(p_n)}{k^2\, q_n^2\, p_n^2}
  \nn\\
  &&\hspace*{2cm}+ \frac{g^2\, N_c\, Z_1}{3} \0{1}{L^4}\sum_{n}
  \frac{H_{3g}(p_n,q_n,k)\,
    N(p_n,q_n,k)\, Z_L(q_n)\, Z_L(p_n)}{k^2\, q_n^2\, p_n^2} \; ,\\
  \frac{1}{G_L(k)} & =& \tilde Z_3 -g^2 \, N_c \, \widetilde Z_1
  \0{1}{L^4}\sum_{n}\frac{k \cP(p_n) q_n\,\, G_L(q_n)\,
    Z_L(p_n)}{k^2\, q_n^2\, p_n^2} \; , \label{eq:ghost}
\end{eqnarray} 
with $p_n=k-q_n$. We furthermore use the translational invariance of the
sums under $q_n\to q_n+2\pi m/L$ with $m\in \ZZ^4$. The abbreviation
$k \cP(p) q$ denotes a contraction with the transverse momentum
tensor, {\it i.e.}  $k \cP(p) q = k_\mu P_{\mu \nu}(p) q_\nu$ with
$P_{\mu\nu}(p)=\delta_{\mu\nu}-p_\mu p_\nu /p^2$.  The integration
kernel $N(p,q,k)$ and the Ansatz for the dressing function
$H_{3g}(p,q,k)$ of the three-gluon vertex can be found in
Refs.~\cite{Fischer,Fischer:2005ui}.  Note also that the above
equations are not restricted to external momenta $k_n = (2\pi/L)\, n
$, and $k$ may or may not be one of these discrete values. For $k\neq
(2\pi/L)\, n $, the Equations \eq{eq:gluon}, \eq{eq:ghost} should be
symmetrised in $q_n$ and $p_n$.
 
The renormalisation is done in the infinite volume limit, which
suffices for finiteness, analogously to finite temperature field
theory. This can be shown on the level of the functional DSEs by
adding and subtracting the infinite volume DSEs and identifying the
renormalisation constants $Z_i \in \{Z_1, \widetilde{Z}_1, Z_3,
\widetilde{Z}_3\}$ at infinite volume with those in a finite volume
$Z_{i,L}$,
\begin{eqnarray}\label{eq:relRG} 
Z_{i,L}=Z_i\,.
\end{eqnarray}
Such a scheme facilitates the approach towards infinite volume. We
emphasise that the scheme does not imply the same RG conditions at
finite and infinite volume. Applying the renormalisation scheme
\eq{eq:relRG} to the propagator DSEs \eq{eq:ghost} and \eq{eq:gluon},
and furthermore using the result $\widetilde Z_1=1$ in Landau gauge
\cite{Taylor:1971ff}, we obtain
 \begin{eqnarray} 
 \hspace*{-6mm}
      \label{eq:gluonren} 
      \frac{1}{Z_L(k)} &=& \0{1}{Z(k'^2)}-\frac{g^2 \, N_c}{3}
      \left(\0{1}{L^4}\sum_{n} \frac{q_n \cP(k) p_n\,\, G_L(q_n)\, 
          G_L(p_n)}{
          k^2\,q_n^2\, p_n^2} 
        -\int \0{d^4 q}{(2\pi)^4} \frac{q \cP(k') p'\,\,
          G(q)\, G(p')}{k'^2\, q^2\, p'^2}\right) \nonumber \\
      &&\hspace*{11mm} + \frac{g^2\, N_c\, Z_1}{3} \left(\0{1}{L^4}\sum_{n}
        \frac{H_{3g}(p_n,q_n,k)\, N(p_n,q_n,k)\, Z_L(q_n)\, 
          Z_L(p_n)}{k^2\, q_n^2\,p_n^2}
      \right.\nonumber\\
      &&\hspace*{50mm}  - \left. \int \0{d^4 q}{(2\pi)^4} 
        \frac{H_{3g}(p',q,k')\,
          N(p',q,k')\, Z(q)\, Z(p')}{k'^2\, q^2\, p'^2}\right),
   \end{eqnarray} 
   \vspace*{-9mm}
 \begin{eqnarray} \label{eq:ghostren}
 \hspace*{-6mm}
      \frac{1}{G_L(k)} &=& \0{1}{G(k'^2)} 
         - g^2 \, N_c \,  \left( \0{1}{L^4}\sum_{n}
	   \frac{k \cP(p_n) q_n\,\, G_L(q_n)\, 
Z_L(p_n)}{k^2\, q_n^2\, p_n^2} \right.\nonumber\\
	&&\hspace*{47mm} \left.
	-  \int \0{d^4 q}{(2\pi)^4}
        \frac{k' \cP(p') q\, \, G(q)\, Z(p')}{k'^2\, q^2\,
          p'^2}\right) \,. 
 \end{eqnarray}
 The sum of the renormalisation terms in each of the two equations is,
 of course, momentum independent. The dependence on $k'$ (with
 $p'=k'-q$) is thus an illusion. Any momentum value can be used
 equally. A particular and convenient choice below will be $k'=k$ (and
 $p'= p$). 

 The RG-scheme in \eq{eq:gluonren},\eq{eq:ghostren} facilitates our
 discussion of finite volume effects in the next section. The
 subtraction is $O(4)$-symmetric and hence keeps the $O(4)$-symmetry
 violations by the finite volume at a minimum. Moreover, this
 subtraction scheme is not restricted to the ghost and gluon
 propagators. It can be extended self-consistently to the DSEs of
 general Green's functions and general truncation schemes
 \cite{Pawlowski:2005xe}. The implementation of this scheme in our
 numerical treatment of the propagator DSEs is discussed in
 Sec.~\ref{sec:renorm}.

\subsection{Zero modes} \label{sec:sszero}

An important subtlety concerns the ghost propagator and vertex
functions. In the covariant gauges there are constant ghost zero modes
which decouple completely from the theory. This is a consequence of
the unfixed global gauge symmetry. The action does not depend on these
constant ghost/anti-ghost modes, as it only couples to derivatives of
the ghost and anti-ghost fields. As a consequence, the Faddeev-Popov
operator can only be inverted for non-vanishing momenta $p\not=0$. The
ghost propagator is thus only defined with the corresponding
projection such that $G_L(q)$ has the property that 
\begin{eqnarray}\label{eq:ghostdecouple} 
G_L(q_n)=G_L(q_n)(1-\delta_{n0})\, .
\end{eqnarray} 
\Eq{eq:ghostdecouple} entails that the zero ghost-momentum terms are
absent in the propagator DSEs. Note that in contrast the gluon zero
modes do contribute to the loop sums as the gluonic vertex functions
do depend on the constant gauge field. However, transversality is not
well-defined at vanishing gluon momentum.  Hence, we have to separate
the zero-momentum contribution explicitly, writing,
\begin{equation}
  D_{L\, \mu\nu}(p_n) =  \delta_{\mu\nu} \, {L^2} \, {C_{L}}
  \,\delta_{n0}\, +\,
    P_{\mu\nu}(p_n)\, \frac{Z_L(p_n)}{p_n^2} \, (1 - \delta_{n0}) \; .
\label{glp_zerosep}
\end{equation} 
Here $C_L$ is some dimensionless constant which represents the value
of the gluon propagator at momentum zero in units of $L^2$. In
principle, this constant is to be determined self-consistently from
the set of algebraic equations representing the propagator DSEs for
the discrete momenta in the periodic $L^4$ box along with the
remaining $Z(p_n)$.  Up to the factor $4/3$ from the tensor structure,
the value of the gluon propagator at momentum zero in the finite
volume is expected to be of approximately the same magnitude as for
the lowest non-zero modes $p_n$. This is what is typically observed in
lattice simulations also. For the parametrisation (\ref{glp_zerosep})
it implies that
\begin{equation}
C_L \sim \frac{Z_L(2\pi/L)}{(2\pi)^2} \; . \label{zeroest}
\end{equation}
Eqs.~(\ref{eq:ghostdecouple}) and (\ref{glp_zerosep}) are important to
obtain well-defined algebraic equations from the DSEs. These equations
are closed among themselves only when the external momentum coincides
with one of the discrete momenta $k = k_m = (2 \pi/L) \, m$ for $m\in
\ZZ^4$.  With Eqs. \eq{eq:ghostdecouple} and (\ref{glp_zerosep}) we
rewrite the gluon DSE \eq{eq:gluon} for $Z(k_m)$ with $m\not = n$ and
$p_{mn} = k_m - q_n$, and separately for $C_L$ with $m=0$, as
\begin{eqnarray}\label{eq:DSEgluon}
  \frac{1}{Z_L(k_m)} &=& Z_3 \, +\,
\frac{g^2N_c}{3 \, L^4}\sum_{n\neq 0,\, m}
  \frac{k_m^2q_n^2-(k_m q_n)^2}{k_m^4 q_n^2
    p_{mn}^2}\, G_L(q_n)G_L(p_{mn}) \, + \, 
  \cdots \\
&=& Z_3 \, +\,
\frac{g^2N_c}{3} \,\frac{1}{(2\pi)^4}\sum_{n\neq 0,\, m}
  \frac{m^2 n^2-(m n)^2}{m^4 n^2
    (m-n)^2} \, G_L(q_n)G_L(p_{mn}) \, + \, \cdots \; , \;\; m\not=0\;,
 \nonumber \\
C_L^{-1} &=& \frac{g^2 N_c}{4} \, \frac{1}{L^2} \sum_{n\not= 0} 
   \frac{1}{q_{n}^2} \, G_L^2(q_n) \, + \, \cdots \, = \, 
 \frac{g^2 N_c}{4} \, \frac{1}{(2\pi)^2} \sum_{n\not= 0} 
   \frac{G_L^2(q_n)}{{n}^2} \,   + \, \cdots   \; .  \label{DSEzerogluon}
\end{eqnarray}
We did not repeat the contributions from the 3-gluon loop explicitly
here again. Note that in the gauge-invariantly regularised full theory
gauge invariance implies that the DSE (\ref{DSEzerogluon}) for $C_L$
must be ultraviolet finite, and of the order given by (\ref{zeroest}).
Hence there is no counter-term in the equation for $C_L$.  The term
from the ghost loop alone, which is given explicitly here, is not
finite, of course. Nonetheless, its apparent ultraviolet divergence,
which has to be cancelled by those of the terms not given explicitly
here, does not affect the infrared analysis, as will be shown below.

With the separation (\ref{glp_zerosep}), the ghost DSE
(\ref{eq:ghost}) for $m\not= 0$ becomes,  
\begin{eqnarray}
 \frac{1}{G_L(k_m)} &=& \widetilde Z_3 -\frac{g^2N_c}{L^4}
  \sum_{n\neq 0,\, m}\frac{k_m^2 q_n^2-(k_m q_n)^2}{k_m^2 q_n^2
  p_{mn}^4} G_L(q_n)  Z_L(p_n)  
 \, - \,  \frac{g^2 N_c\, C_L}{L^2k_m^2}   \, G_L(k_m) \; . 
 \label{eq:DSEghost} \end{eqnarray} 
There is no DSE for $m=0$ in this case. In the momentum range
(\ref{eq:valrange}) of interest here, we have $(2\pi)^2 \ll L^2k_m^2 $.
With dressing functions $G(k_m) $, $Z(k_m) $ of the order
one, and $C_L$ of the order of magnitude given by (\ref{zeroest}), the
gluonic zero mode contribution to the ghost DSE (\ref{eq:DSEghost}) is
therefore suppressed by an explicit factor of $(2\pi)^2/(L^2k_m^2)$ as
compared to the other terms in this momentum range. 

An important practical consequence of the above analysis is that we
can safely drop the contributions of the gluonic zero mode in the
DSEs.  This is important for our numerical analysis, {\it c.f.},
Sec.~\ref{sec:snum}.

\subsection{Infrared analysis} \label{sec:ssvols}

We proceed with an infrared analysis following that of
\cite{Lerche:2002ep,Pawlowski:2003hq}. Due to the missing
$O(4)$-symmetry the dressing functions $Z_L(k)$ and $G_L(k)$ are
functions of the individual components $k_i$, and not of $k^2$ alone.
When using an $O(4)$-symmetric ultraviolet cut-off in
\eq{eq:gluon},\eq{eq:ghost}, however, we have approximate 
$O(4)$ symmetry for momenta $k^2 L^2\geq 1$.  Therefore, in order
to simplify the analysis, we may neglect the residual $O(4)$-symmetry
violation considering an $O(4)$-average with respect to the external
momentum $k$ in the propagator DSEs.

For $k\to 0$, with $k\neq k_m = (2 \pi/L)\, m$, the gluon equation
\eq{eq:DSEgluon} only depends on $G_L(q_n)$ and $G_L(p_n) \to
G_L(q_n)$ (where $p_n = k - q_n $, {\it c.f.},
Sec.~\ref{sec:torusDSEs}, and thus $p_n\to q_n $) with $q_n^2 \geq
4\pi/L^2$.  Note that this momentum regime is below the lowest
momentum obtainable in the numerical calculations in
Sec.~\ref{sec:snum}.

The ghost dressing functions in the ghost loop of (\ref{eq:DSEgluon})
therefore remain finite in this limit (for monotonically decreasing
$G_L(q_n)$ the upper bound will be $G_L(2\pi/L)$). Hence, we deduce
from the right hand side of the gluon equation that $Z(k)/k^2$ has a
finite yet direction-dependent limit for $k\to 0$. The angle dependence
in this limit relates to the missing $O(4)$-symmetry on the torus.
After angular averaging we thus write
\begin{eqnarray}\label{eq:Zir}
  Z_L(k^2) \equiv \int \0{d\omega_k}{2\pi^2} \,  
  Z_L(k)  \, \propto \, \0{k^2}{\mu^2_Z(L)}\left(1+\CO(k^2/\mu_Z^2)\right)\,,
\end{eqnarray}
for $k\to 0$. \Eq{eq:Zir} displays 
the behaviour of a dressing function of a particle with a screening
mass  $\mu_Z(L)$. Inserting this behaviour into the ghost equation
(\ref{eq:DSEghost}) we analogously deduce for the ghost dressing
function that 
\begin{eqnarray}\label{eq:Gir}
  G_L(k^2) \equiv \int\0{d\omega_k}{2\pi^2} \,  G_L(k) \, \propto \, 1
   +\CO(k^2/\mu^2_G)\,, 
\end{eqnarray}
for $k\to 0$. In contrast to the gluon propagator, the ghost
propagator remains massless in the finite volume. Due to the ghost
screening mass $\mu_G(L)$ the ghost dressing function $G_L(k)$ is
infrared finite at finite $L$, however. The infinite-volume
renormalisation scheme of Eqs.~\eq{eq:relRG}, \eq{eq:gluonren} and
\eq{eq:ghostren} will guarantee that both screening masses
$\mu_{Z/G}(L)\to 0$ in the infinite volume limit $L\to\infty$.

By construction, the averaged dressing functions only depend on the
magnitude of momentum. We emphasise that this approximation does not
change powers in $k^2$. In particular, it will not affect the infrared
exponents of momenta $k\gg 2\pi/L$. In the regime \eq{eq:valrange} we
can expand the dressing functions $Z_L(k^2)$ and $G_L(k^2)$ about
their leading physical infrared behaviour,
\begin{eqnarray}\nonumber
  Z_L(k^2)& = & Z_\mathrm{IR}\left(\0{\Lambda^2_{\rm QCD}}{
k^2+\mu_{Z}^2}\right)^{\kappa_{Z_L}}
\, \0{k^2}{k^2+\mu_{Z}^2} 
  \left(1+\delta Z\left(
    \0{k^2}{\Lambda^2_{\rm QCD}}, 
    \0{1}{k^2 L^2}\right)\right)\,,\\ 
  G_L(k^2)& = & G_\mathrm{IR}
\left(\0{\Lambda^2_{\rm QCD}}{ k^2+\mu_G^2}\right)^{\kappa_{G_L}} \,
  \left(1
    +\delta G\left(\0{k^2}{\Lambda^2_{\rm QCD}}, 
    \0{1}{k^2 L^2}\right)\right)\,, 
 \label{eq:expand}\end{eqnarray} 
with
\begin{eqnarray}\label{eq:0}
\delta Z(0,0)=\delta G(0,0)=0\,, 
\end{eqnarray} 
and possibly $k$-dependent $\kappa_{Z_L/G_L}$ but constant
$Z_\mathrm{IR}$, $G_\mathrm{IR}$.  For $k\to 0$ we arrive at the
limits \eq{eq:Zir} and \eq{eq:Gir} with subleading powers depending on
$k^2/\mu^2_{Z/G}$.  The finite volume corrections are suppressed with
$1/( k^2 L^2)$, and the confinement scale works as an UV cut-off
leading to corrections of the order $k^2/\Lambda^2_{\rm QCD}$.  The
corresponding terms are hidden in the corrections $\delta Z$ and
$\delta G$ respectively. We could also have absorbed the screening
'masses' $\mu_{Z/G}$ into the definitions of the $\delta Z/\delta
G$-corrections, but we prefer to keep them explicitly in this infrared
Ansatz for later convenience.

For $\Lambda_{\rm QCD}^2\gg k^2\gg \mu_{G/Z}^2$ we are left with the
physical infrared behaviour, $Z_L(k^2)\sim (k^2)^{-\kappa_{Z_L}}$,
$G_L(k^2)\sim (k^2)^{-\kappa_{G_L}}$. In particular, for $L\to\infty $
the parametrisations (\ref{eq:expand}) tend towards the infinite volume 
infrared forms,
\begin{eqnarray}\nonumber
  Z_L(k^2) \to Z(k^2) & =& Z_\mathrm{IR}\left(\0{\Lambda^2_{\rm QCD}}{
k^2}\right)^{-2\kappa }
\, \left(1+\delta Z\left(
    \0{k^2}{\Lambda^2_{\rm QCD}}, 
    0 \right)\right)\,,\\ 
  G_L(k^2)\to G(k^2) & = & G_\mathrm{IR}
\left(\0{\Lambda^2_{\rm QCD}}{ k^2 }\right)^{\kappa} \,
  \left(1
    +\delta G\left(\0{k^2}{\Lambda^2_{\rm QCD}}, 
    0 \right)\right)\, ,
 \label{eq:expand-asymp}\end{eqnarray} 
with the limits
\begin{equation}
 \kappa_{Z_L} \to -2\kappa\; , \;\; \kappa_{G_L} \to \kappa\; , \; \;
\mbox{and} \;\; \mu_{Z/G} \to 0 \;\;  \mbox{for} \;\; 
L\to \infty \; .
\end{equation} 
This follows immediately from the renormalised DSEs
\eq{eq:gluonren}, \eq{eq:ghostren}.

We are also quantitatively interested in the approach towards the
infinite volume solutions, and concentrate on momenta
\begin{eqnarray}\label{eq:deepIR} 
\0{k^2}{\Lambda^2_{\rm QCD}}\to 0\,,
\end{eqnarray} 
which is achieved formally by taking the limit $\Lambda^2_{\rm QCD}
\to\infty$. The limit of asymptotically large volumes, on the other
hand, is obtained from $k^2\to \infty$.  In general, the
infinite-volume DSEs will require renormalisation also for
$\Lambda^2_{\rm QCD} \to\infty$.  This is different from the standard
perturbative UV-renormalisation in that contributions of momenta at
about $\Lambda_{\rm QCD}$ are removed. Once the infinite volume DSEs
are properly renormalised, however, the limit $k^2\to \infty$ can be
performed without further subtractions in the torus DSEs \eq{eq:gluon}
and \eq{eq:ghost}: when the right hand sides in the form of
\eq{eq:gluonren} and \eq{eq:ghostren} are used, the results remain
finite.

We first insert the parametrisations \eq{eq:expand} into the gluon
Dyson-Schwinger equation \eq{eq:gluonren}. After angular averaging, we
consider the leading contribution from the ghost loop in the momentum
regime (\ref{eq:valrange}) which becomes
\begin{eqnarray}
  \frac{1}{Z_\mathrm{IR}}\, \left(\0{\Lambda^2_{\rm QCD}}{
      k^2+\mu_{Z}^2}\right)^{-\kappa_{Z_L}}
  \, \0{k^2+\mu_{Z}^2}{k^2} & = &  \frac{g^2 N_c}{3} \,
  \frac{G_\mathrm{IR}^2} {L^4} \,\times   \label{IRghostloop}\\
  && \hskip -.2cm \sum_{n\neq 0,\, n_k}
  \frac{q_n \CP(k) q_n}{k^2 q_n^2
    p_{n}^2}\,
  \left(\frac{\Lambda_\mathrm{QCD}^2}{q_n^2+\mu_G^2}
  \right)^{\kappa_{G_L}}
  \left(\frac{\Lambda_\mathrm{QCD}^2}{p_n^2+\mu_G^2}
  \right)^{\kappa_{G_L}}
  + \, 
  \cdots \; .  \nonumber
\end{eqnarray}
Herein, $n_k$ labels the loop momentum closest to $k$, {\it i.e.}, the
minimum of $p_n^2$. For momenta in the range (\ref{eq:valrange}) the
corrections due to $\delta Z$, $\delta G$ can be neglected. Moreover,
because the ghost loop dominates in this momentum regime, we can
safely drop all the other contributions, not given explicitly here. 
The infinite volume renormalisation vanishes due to the value of
$\kappa_c>0.5$ in \eq{kappa_c}: the momentum integral in \eq{eq:gluonren} 
is finite and agrees with $1/Z(k'^2)$. We immediately conclude that 
\begin{eqnarray}\label{eq:k-limit}
\lim_{q^2\to\infty}\kappa_{G_L}(q^2) > 0.5\,, \end{eqnarray}
to guarantee finiteness of \eq{IRghostloop}. 
In the approximate expression (\ref{IRghostloop}),
$\Lambda_\mathrm{QCD}^2/k^2 $ and $ k^2/\mu_{Z/G}^2$ are all assumed
to be sufficiently large but still finite. When these ratios tend to
infinity we recover one of the infinite volume DSE conditions for the
critical exponent $\kappa_c$ (\ref{kappa_c}) and the infrared fixed
point $\alpha_c$ (\ref{IRfixed}) in \cite{Lerche:2002ep}. 

Note also that the momentum sums in \eq{IRghostloop} receive their main 
contribution for momenta at about $k^2$. Upon rearranging (\ref{IRghostloop}) 
we obtain up to sub-leading terms, 
\begin{equation} 
  \0{4 \pi}{\alpha_c N_c } =
  \left(\0{\Lambda_\mathrm{QCD}^2}{k^2+\mu^2_Z}\right)^{\kappa_{Z_L}+
  2\kappa_{G_L} }\,   I_{Z_L}(\kappa_{G_L},k,\0{2\pi}{L}) \; , 
  \label{gluonlimit}
\end{equation}
where
\begin{equation}
  I_{Z_L}(\kappa, k,\0{2\pi}{L}) =
  \0{16\pi^2}{3L^4} \sum_{n\not= 0, n_k} \0{1}{ p_n^2 q_n^2} \,
\0{q_n  \CP(k) q_n}{k^2+\mu^2_Z} 
  \,  \left(\0{k^2+\mu^2_Z}{q_n^2+\mu^2_G}\right)^{\kappa} 
  \left(
    \0{k^2+\mu^2_Z}{p_n^2+\mu^2_G}\right)^{\kappa}  \,, \label{IZL}
\end{equation}
and $\alpha_c = \big(g^2/4\pi\big)\,  Z_\mathrm{IR}
G_\mathrm{IR}^2$. 
Recall that Eq.~(\ref{gluonlimit}) holds for $\mu_{Z/G}^2 \ll k^2 \ll
\Lambda_\mathrm{QCD}^2$. For its left hand side to have a
non-vanishing and finite value in the limit (\ref{eq:deepIR}) with
$\Lambda_\mathrm{QCD}^2 \to\infty$,  we need
to have at sufficiently large but still finite $L$, as in the infinite
volume limit,  
\begin{equation} \label{kapparel}
   \kappa_{Z_L} + 2 \kappa_{G_L} = 0\; . 
\end{equation}
Thus we define analogously,
\begin{equation} 
    \kappa_{G_L} \equiv  \kappa_{L} \; ,\;\;
    \kappa_{Z_L} \equiv - 2\kappa_{L} \; ,
\end{equation}
as in the $L\to\infty$ limit. 
Then, the dependence on the scale $\Lambda_\mathrm{QCD}$ is eliminated 
from the infrared analysis as necessary. We obtain,
\begin{equation} 
  \0{4 \pi}{\alpha_c N_c } =  I_{Z_L}(\kappa_{L},k,\0{2\pi}{L}) \; , 
  \label{gluonlimit2}
\end{equation}
and it is easy to verify that in the second limit, that of large volume via
$k^2 \to\infty$, this explicitly leads to the 
infrared condition (\ref{IRfixed}) for $\alpha_c$ from the
infinite-volume gluon DSE in \cite{Lerche:2002ep}, as   
\begin{eqnarray} 
 I_{Z_L}(\kappa_{L},k,\0{2\pi}{L}) \to I_Z(\kappa_L)  &=& 16\pi^2 \int
 \frac{d^4q}{(2\pi)^4}\,  \frac{1}{q^2 p^2} \,
\frac{q\CP(k) q}{3 k^2} 
 \left(\frac{k^2}{q^2}\right)^{\kappa_L} \, 
\left(\frac{k^2}{p^2}\right)^{\kappa_L} \label{IZcont}\\ 
 &=& \frac{1}{2} \, \frac
 {\Gamma^2(-\kappa_L) \, \Gamma(2\kappa_L-1)}{\Gamma^2(\kappa_L-1) \,
 \Gamma(4-2\kappa_L) }  \; , \nonumber
\end{eqnarray}
for $k^2\to\infty $. The condition in Eq.~(\ref{gluonlimit2})
therefore implies that $\kappa_L \!\to\! \kappa_c $ in this limit, as
expected.

The infrared dominant contribution for
$\Lambda_\mathrm{QCD}^2\to\infty $ in the ghost DSE still needs
ultraviolet renormalisation, as it does in the infinite volume case.
With (\ref{eq:expand}), (\ref{eq:expand-asymp}) we obtain the
corresponding contributions to the renormalised ghost DSE
(\ref{eq:ghostren}), with $k'=k$,
\begin{eqnarray}
  \frac{1}{G_\mathrm{IR}}\, \left(\0{\Lambda^2_{\rm QCD}}{
      k^2+\mu_{G}^2}\right)^{-\kappa_{L}} \!\!
  & = &    \frac{1}{G_\mathrm{IR}}\, \left(\0{\Lambda^2_{\rm QCD}}{
      k^2}\right)^{-\kappa_c} \!\!\! -   {g^2 N_c}
      G_\mathrm{IR}Z_\mathrm{IR} \, \Bigg(
  \frac{1} {L^4} \sum_{n\neq 0,\, n_k}
  \frac{k \CP(p_n) k}{k^2 q_n^2
    (p_{n}^2 + \mu_Z^2)}\, \times     \label{IRghostDSE}\\
  && \hskip -1.6cm 
  \left(\frac{\Lambda_\mathrm{QCD}^2}{q_n^2+\mu_G^2}
  \right)^{\kappa_{L}}
  \left(\frac{\Lambda_\mathrm{QCD}^2}{p_n^2+\mu_Z^2}
  \right)^{-2\kappa_{L}}\!
  - \, \int \frac{d^4q}{(2\pi)^4} \frac{k \CP(p) k}{k^2 q^2 p^2} 
\left(\frac{\Lambda_\mathrm{QCD}^2}{q^2}
  \right)^{\kappa_{c}}
  \left(\frac{\Lambda_\mathrm{QCD}^2}{p^2}
  \right)^{-2\kappa_{c}}\Bigg)
  \, .  \nonumber
\end{eqnarray}
This equation is ultraviolet finite because the (weakly) 
momentum dependent exponent $\kappa_L \!\to\! \kappa_c $, whenever the
momentum argument of the corresponding dressing function becomes large (of the
order of $\Lambda_\mathrm{QCD}$), as already derived from the ghost loop in
the gluon DSE above. Eq.~(\ref{IRghostDSE}) can be rewritten in the form,
\begin{equation} 
  \0{4 \pi}{\alpha N_c }
  \,=\, I_{G_L}(\kappa_L,k,\0{2\pi}{L})-\left(\0{k^2}{k^2+
        \mu^2_G}\right)^{\kappa_L} 
\left(\0{\Lambda^2_\mathrm{QCD}}{k^2}\right)^{\kappa_L-\kappa_c}
\left[I_G(\kappa_c,k,0)-I_{G}(\kappa_c)\right] \; . \label{ghost-subtract}
\end{equation}
Here, $I_G(\kappa,k,2\pi/L)$ is the corresponding loop sum (as
explicitly given in (\ref{IGL}) below), which is ultraviolet divergent
in this case.  This divergence is cancelled by the likewise
ultraviolet divergent infinite volume integral expression
\begin{equation}
  I_G(\kappa, k,0) \,=\,
  - 16\pi^2 \int \0{d^4 q}{(2 \pi)^4}  \left(\0{1}{q^2}\right)^2
  \left(\0{q^2}{k^2}\right)^{1-\kappa} \left(
    \0{k^2}{p^2}\right)^{1-2\kappa} \0{k_\mu
  \CP_{\mu\nu}(p) k_\nu}{k^2} \; , 
 \label{eq:Icont} 
\end{equation}
whose finite contributions  are independent of $k$
and given by \cite{Lerche:2002ep}
\begin{equation} 
  I_G(\kappa)= -\0{3}{2}\0{\Gamma^2(-\kappa)
    \Gamma(2 \kappa-1)}{
    \Gamma(-2\kappa)\Gamma(\kappa-1)\Gamma(\kappa+3)}\; .
\label{IGcont}
\end{equation}
That leaves only the divergent part in the terms in brackets on the
right of Eq.~(\ref{ghost-subtract}) to remove the ultraviolet
divergence of the loop $I_G(\kappa_L,k,2\pi/L)$ (with
$\kappa_L\!\to\!\kappa_c$ at large
momenta). Eq.~(\ref{ghost-subtract}) is the analogue to Eq.~(\ref{gluonlimit2})
 from the gluon DSE. 

Alternatively,
we can use $k' = 0 $ in the renormalisation terms of 
(\ref{eq:ghostren}) to compute the infinite-volume ghost
renormalisation constant $\widetilde Z_3$ explicitly \cite{Lerche:2002ep},  
\begin{equation}
 \widetilde Z_3 = g^2 N_c \frac{3}{4} \int \frac{d^4q}{(2\pi)^4} \,
 \frac{Z(q^2)G(q^2)}{q^4} \; . 
\end{equation}
Using the infinite volume infrared forms (\ref{eq:expand-asymp}) for
 asymptotically large $\Lambda_\mathrm{QCD}$ with $O(4)$-symmetric
 ultraviolet cut-off $\Lambda_\mathrm{UV}$ this becomes
\begin{equation}
 \widetilde Z_3 \to \widetilde Z_3^\mathrm{IR} = g^2 N_c \frac{3}{4} \,
 \frac{Z_\mathrm{IR}G_\mathrm{IR}}{16\pi^2}
 \left(\frac{\Lambda_\mathrm{UV}^2}{\Lambda_\mathrm{QCD}^2}\right)^\kappa
 \; . \label{UV}
\end{equation}
Of course, asymptotically large $\Lambda_\mathrm{QCD} $ in this case
means $\Lambda_\mathrm{QCD} \to \Lambda_\mathrm{UV}$, and
$\Lambda_\mathrm{QCD} $ acts as the ultraviolet momentum cut-off in
the infrared analysis as explained above. Then, with (\ref{eq:expand})
in the ghost DSE (\ref{eq:ghost}), we obtain from the leading infrared
contribution in this limit
\begin{equation} 
  \0{4 \pi}{\alpha_c N_c }
  = I_{G_L}(\kappa_L,k,\0{2\pi}{L}) + \frac{3}{4} 
   \left(\0{\Lambda_\mathrm{QCD}^2}{k^2+
        \mu^2_G}\right)^{\kappa_L}  \; .
\label{ghostlimit} 
\end{equation}
The loop sum herein is cut off at $q_n^2 \le \Lambda^2_\mathrm{QCD} \equiv
(2\pi/L)^2 N^2_\Lambda$, 
\begin{equation} 
  I_{G_L}(\kappa_L, k, \0{2\pi}{L}) =
  - \frac{16\pi^2}{L^4}\,  \sum^{n^2\leq N_\Lambda^2}_{n\not= 0, n_k}
  \0{k \CP(p_n)
  k}{k^2 q_n^2 (p_n^2+ \mu_Z^2) }
  \left(\0{k^2+\mu^2_G}{q_n^2+\mu^2_G}
  \right)^{\kappa_L}  
  \left(
    \0{p_n^2+\mu^2_Z}{k^2+\mu^2_G}\right)^{2\kappa_L}  \label{IGL}
\end{equation}
where for asymptotically large $q_n^2$, approaching
$\Lambda_\mathrm{QCD}^2 $, the sum over the loop-momentum $q_n$ tends
to the corresponding integral because of the large number $L^4 q_n^2
dq_n^2/(16\pi^2) $ of modes in $[q_n^2,q_n^2+dq_n^2]$. The ultraviolet
divergence of the loop with cut-off $\Lambda_\mathrm{QCD}$ is
therefore subtracted correctly by the infinite volume counter-term in
(\ref{ghostlimit}), which thus remains finite in the
$\Lambda_\mathrm{QCD}^2 \to \infty$ limit as required.

As for the ghost-loop contribution to the gluon DSE, {\it c.f.},
Eq.~(\ref{gluonlimit2}) with (\ref{IZcont}), 
in the second limit of large $L^2k^2 $ we recover the infrared
self-consistency condition from the infinite-volume ghost DSE of
\cite{Lerche:2002ep}. Using Eqs.~(\ref{gluonlimit2}) and
(\ref{ghostlimit}), or (\ref{ghost-subtract}), we furthermore obtain
the leading corrections $\propto 1/(k^2L^2)$ as follows:

First, expanding $I_{Z_L}(\kappa_L,k,2\pi/L)$ in Eq.~(\ref{IZL}) for
large $k^2$, from the ghost loop of the gluon DSE (\ref{gluonlimit2}),
we obtain
\begin{equation}
   \0{4 \pi}{\alpha_c N_c } 
   = I_Z(\kappa_L) \, \left\{ 1 + (2\kappa_L -1 )
  \frac{\mu_Z^2}{k^2} - 2\kappa_L\, \frac{(2\kappa_L -1)
  (3-2\kappa_L)}{(1+\kappa_L)(1-\kappa_L)} \,  \frac{\mu_G^2}{k^2} \right\}
   \, + \, \mathcal{O}\Big(\frac{1}{L^4k^4}\Big) \; .  \label{ZIRL}
\end{equation}
Analogously, from the ghost DSE, condition (\ref{ghostlimit}) becomes
\begin{eqnarray}
  \0{4 \pi}{\alpha_c N_c } 
   &=& I_G(\kappa_L) \, \left\{ 1 - \left( \kappa_L 
   - \frac{\kappa_L^2 \,(2+\kappa_L)}{(1+\kappa_L)(1-\kappa_L)} \right) \,
   \frac{\mu_G^2}{k^2} \right.
                                \label{GIRL} \\
&& \hskip 6cm \left.  - \frac{ (2\kappa_L-1)(2+\kappa_L)}{4(1-\kappa_L)}\,
   \frac{\mu_Z^2}{k^2} \right\} 
      \, + \, \mathcal{O}\Big(\frac{1}{L^4k^4}\Big) \; . \nonumber
\end{eqnarray}
The leading order herein again reproduces the corresponding infinite volume 
condition which entails $\kappa_L \to \kappa_c$ in the limit
$k^2\to\infty$, as
\begin{equation} 
\0{4 \pi}{\alpha_c N_c } 
   = I_Z(\kappa_c) 
   = I_G(\kappa_c) \; .
\end{equation}
With $I_Z(\kappa) $,  $I_G(\kappa) $ from
Eqs.~(\ref{IZcont}), (\ref{IGcont}) it follows that $\kappa_c = 
(93 - \sqrt{1201})/98 = 0.59535... $ and $\alpha_c = 2.9717... $
(for $N_c = 3$), see \cite{Lerche:2002ep}.

Secondly, for the leading corrections to this limit, we furthermore let 
\begin{equation}
\kappa_L\equiv \kappa_L(L^2k^2) = \kappa_c - \frac{c_\kappa}{L^2k^2} + 
 \mathcal{O}\Big(\frac{1}{L^4k^4}\Big) \; . \label{kappaLim}
\end{equation}
In addition, parametrising the leading $L$-dependence of the screening
masses via 
\begin{equation} 
  \mu_Z^2 = \frac{c_Z}{L^2} \; , \;\; \mbox{and} \;\; 
  \mu_G^2 = \frac{c_G}{L^2} \; ,   \label{muLim}
\end{equation}
we obtain from the leading corrections in (\ref{ZIRL}) and (\ref{GIRL}),
respectively,
\begin{eqnarray} 
  c_\kappa \, \frac{I'_Z(\kappa_c)}{I_Z(\kappa_c) } &=& 
(2\kappa_c -1 )
  \, c_Z\,  - \, 
  \frac{\kappa_c(2+\kappa_c)}{6(1-\kappa_c)} \, c_G \; ,\label{L-corr} \\
  c_\kappa \, \frac{I'_G(\kappa_c)}{I_G(\kappa_c) } &=& 
  - \frac{ (2\kappa_c-1)(2+\kappa_c)}{4(1-\kappa_c)}\,
   c_Z
 - \left(\kappa_c -  \frac{\kappa_c^2 \,
   (2+\kappa_c)}{(1+\kappa_c)(1-\kappa_c)}\right) \, c_G 
\; .\nonumber
\end{eqnarray}
The fact that we have 3 unknowns, $c_\kappa$, $c_Z$ and $c_G$, from
only 2 equations here reflects an ambiguity in our
parametrisations (\ref{eq:expand}) of the finite-volume infrared
behaviour of the gluon and ghost dressing functions in the momentum regime  
of (\ref{eq:valrange}). As a parametrisation of the leading corrections to
the infinite-volume infrared forms with originally 2 parameters in
each of the dressing functions, $Z_L(k^2)$ and $G_L(k^2)$,
Eqs.~(\ref{eq:expand}) are somewhat redundant, of course.  
Even after eliminating one of the originally 2 exponents, the residual 
redundancy is manifest here. An  $L^2k^2$ independent exponent is
excluded, however, as $c_\kappa = 0$ implies that also $c_Z = c_G = 0$ for
$\kappa_c = (93 - \sqrt{1201})/98 $.  

In fact, in units of $c_\kappa$, Eqs.~(\ref{L-corr}) yield numerical
values
\begin{equation} 
 c_Z \approx 238\, c_\kappa\; , \;\; c_G \approx 86 \, c_\kappa \; .
\label{c-rel}
\end{equation}
Because $c_\kappa >0$ for positive $c_{Z/G}$ ({\it c.f.},
Eqs.~(\ref{muLim})), with the definition of $c_\kappa$ in
(\ref{kappaLim}), we find that $\kappa_L$ approaches $\kappa_c$ from
below. 

Moreover,  Eqs.~(\ref{c-rel}) imply that the momentum dependence of
the exponent $\kappa_L(L^2k^2)$ is a rather weak effect in comparison to
the influence of the screening masses, which thus account for the
dominant finite-volume corrections at  sufficiently large $k^2L^2$. 
Therefore, a quantitatively good approximation for large volumes and
momenta in (\ref{eq:valrange}) is given by 
\begin{eqnarray}\nonumber
  Z_L(k^2)& = & Z_\mathrm{IR}\left(\0{\Lambda^2_{\rm QCD}}{
k^2+\mu_{Z}^2}\right)^{-2 \kappa_{c}}
\, \0{k^2}{k^2+\mu_{Z}^2}\,,\\ 
  G_L(k^2)& = & G_\mathrm{IR}
\left(\0{\Lambda^2_{\rm QCD}}{ k^2+\mu_G^2}\right)^{\kappa_{c}} \,, 
 \label{eq:IRfit}\end{eqnarray}
with the infinite volume $\kappa$'s. With this approximation of
neglecting the momentum dependence of the infrared exponent $\kappa_L$,
estimates of the screening masses $\mu_{Z/G}$ can in principle be
obtained from Eqs.~(\ref{gluonlimit2}) and (\ref{ghost-subtract}),
(or (\ref{ghostlimit})) at $k^2 = 0$ rather than considering large
$k^2$ for the leading $1/(L^2k^2)$ corrections as done so far. Because
the resulting equations will then 
necessarily involve the summations over the discrete loop momenta,
such an estimate would already need a numerical solution to these
equations. Here, we rather present our full numerical solutions to the 
torus DSEs in the next section. The masses extracted from these full
solutions in Sec.~\ref{sec:asym} are in good agreement with the
ratio $c_Z/c_G$ determined here.
 
Finally, inserting the screening masses $\mu_{Z/G} = c_{Z/G}/L^2$ with
$L$-independent coefficients $c_{Z/G}$ into the IR-asymptotics
(\ref{eq:IRfit}), for vanishing momentum $k^2$ with $\delta Z=\delta
G=0$ and $-\kappa_{Z_L}/2 = \kappa_{G_L}  \approx \kappa_c$ in
\eq{eq:expand}, we obtain an estimated behaviour with $L$ of the
infrared limit of the  gluon 
and ghost dressing functions as
\begin{eqnarray} \label{eq:IRrunning} 
\lim_{k^2\to 0} \, \frac{Z_L(k^2)}{k^2}\, &\propto& \, L^{2(1-2\kappa_c)}\;,\\
\lim_{k^2\to 0} \, G_L(0)\, &\propto &\, L^{2\kappa_c}\; .
\end{eqnarray} 
The $L$-dependence of the dressing functions at $k=0$ mimics the
infrared leading momentum behaviour. In particular, as $(1-2\kappa_c)
\approx - 0.19$, this implies that the zero momentum gluon propagator
is expected to decrease extremely slowly with the volume $V=L^4$ towards zero
in the infinite volume limit $\propto V^{-\epsilon}$ with an exponent
$\epsilon = (2\kappa_c-1)/2 $ which is smaller than 0.1. The infrared limit
of the ghost propagator, on the other hand, increases faster than
linear with the length $L$. We will compare these qualitative features to
the full numerical solution in the next section.

\section{Numerical results}\label{sec:snum}

\subsection{The momentum lattice \label{sec:mom}}

\begin{figure}[t!]
\centerline{ 
\epsfig{file=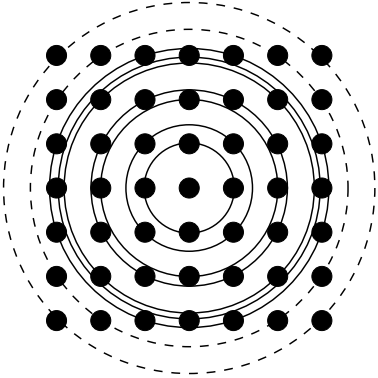,width=50mm}} 
\caption{Sketch of the momentum grid dual to the four-torus. The
  summation over complete hyperspheres is indicated by fully drawn
  circles. The hyperspheres depicted by dashed lines are not complete
  if one uses Cartesian cutoffs instead of an $O(4)$ invariant
  one.}\label{fig:sketch}  

\vspace*{.8cm}

\end{figure}

As already mentioned in the previous section, in a 4-dimensional
hypercubic volume $V=L^4$ with periodic boundary conditions, the
momentum integrals of the infinite volume DSEs are replaced
by sums, {\it c.f.}, \pref{eq:torusints}.  For the numerical treatment
of the equations it is convenient to rearrange these summations in a
spherical coordinate system \cite{Fischer}. A two-dimensional
illustration of this procedure is given in Fig.~\ref{fig:sketch}. We
thus write (\ref{eq:torusints}) in the form
\begin{equation}
\frac{1}{L^4} \sum_{\mbox{\footnotesize $n \in  \ZZ^4 $}}\, (\dots)
=  \frac{1}{L^4} \sum_{\mbox{\footnotesize $m,j$}}\, (\dots) \;,
 \label{sum2}
\end{equation}
where $m$ labels sums over hyperspheres, each containing all momentum vectors
of the same absolute value and $j$ numbers the individual vectors on a given 
hypersphere. The resulting double sum corresponds to the splitting
between radial and angular integrals in the infinite volume
DSEs. This correspondence is very good for large hyperspheres ({\it
  i.e.}, in the ultraviolet momentum region), where a large number of
vectors on a given sphere samples the angular dependence of the
integral kernels well. However, in the infrared, {\it i.e.}, on the innermost 
spheres, this sampling will be poor, thus resulting in hypercubic artefacts.
We will identify these in our solutions in Sec.~\ref{sec:props}.
Note, however, that the sole dependence of the dressing functions on the
squared momenta is not touched by these artefacts. We explicitly verified that
the dressing functions $Z(p^2)$ and $G(p^2)$ do not depend on the direction of 
the momentum $p$. 

For numerical reasons we also have to introduce a momentum cut-off
that limits the extent of the momentum lattice. In the infinite
volume formulation such a cut-off $\Lambda_\mathrm{UV}$ is
O(4)-invariant. The resulting renormalised dressing functions are
independent of the value of $\Lambda_\mathrm{UV}$, as they should be
in a renormalisable quantum field theory. In general, this is no
longer true on a four-torus and we expect cut-off artefacts. It turns
out that these effects are sizeable only when O(4)-symmetry is badly
broken as is the case, {\it e.g.}, when restricting the sums in
(\ref{eq:torusints}) in each Cartesian direction. Much better results
are obtained, if an (approximately) O(4)-invariant cut-off is
introduced by using the summation procedure (\ref{sum2}) and summing
only over such hyperspheres which do not receive any further vectors
when enlarging the momentum grid. In Fig.~\ref{fig:sketch} these are
indicated by the solid lines as opposed to the dashed ones. Since
these fully occupied outer spheres typically contain a large number of
momentum vectors, the corresponding 'angular sum' over $j$ is close in
value to the corresponding angular integrals in the infinite
volume limit and is thus a good approximate representation
of the corresponding O(4)-symmetry. Note that we have used a similar
procedure in our infrared analysis around and below Eq.~(\ref{UV}).

\subsection{Renormalisation procedure \label{sec:renorm}}

Further details of our numerical method to solve the coupled system of
Eqs.~(\ref{eq:ghost}), (\ref{eq:gluon}) have been described in
\cite{Fischer,Fischer:2005ui,Fischer:2003zc} and shall not be repeated
here. The only major difference to previous calculations of DSEs on
the torus is a modification of the renormalisation conditions. These
modifications turn out to have a substantial impact on the large
volume limit.

The propagator DSEs (\ref{eq:gluonren}) and (\ref{eq:ghostren})
resulting from the renormalisation condition (\ref{eq:relRG}) in
Sec.~\ref{sec:sana} make the approach towards the infinite 
volume solutions at large torus volumes transparent. A
direct numerical implementation of eqs.~(\ref{eq:gluonren}) and
(\ref{eq:ghostren}) requires a sufficiently large UV-cutoff
$\Lambda_\mathrm{UV}$. Unfortunately, for reasons of CPU-time we are
restricted to a rather low UV-cutoff of the order of 
$\Lambda_\mathrm{UV}=2-3$ GeV. For such a low cutoff one encounters 
convergence problems in the infinite volume DSEs which prevent a reliable 
extraction of the ghost and gluon renormalisation factors 
$\widetilde{Z}_3$ and $Z_3$. We therefore adjust the renormalisation 
procedure of Sec.~\ref{sec:sana} to the numerics as explained in the 
following.

We substitute the infinite volume subtractions in
eqs.~(\ref{eq:gluonren}) and (\ref{eq:ghostren}) by subtractions on
the torus at a fixed scale $s^2$, which in general does not have to be
equal to the renormalisation point.  In the symbolical notation
\begin{eqnarray}
  \frac{1}{G(p^2)} &=& \widetilde{Z}_3 + g^2 N_c \Pi_{ghost}(p^2) \,, \\
  \frac{1}{Z(p^2)} &=& Z_3 + g^2 N_c \Pi_{glue}(p^2) \,, 
\end{eqnarray} 
for the
equations (\ref{eq:gluon}) and (\ref{eq:ghost}) this procedure yields
\begin{eqnarray}
  \frac{1}{G(p^2)} &=& \frac{1}{G(s^2)} + g^2 N_c \left(\Pi_{ghost}(p^2)
    - \Pi_{ghost}(s^2)\right) \,, \\
  \frac{1}{Z(p^2)} &=& \frac{1}{Z(s^2)} + g^2 N_c \left(\Pi_{glue}(p^2)
    - \Pi_{ghost}(s^2)\right),   
\end{eqnarray}
similarly to (\ref{eq:gluonren}) and (\ref{eq:ghostren}).  The
renormalisation constants $Z_3$ and $\widetilde{Z}_3$ are traded for
two input values at $G(s^2)$ and $Z(s^2)$. These values are
not independent of each other. In the infinite volume 
limit this can be seen from the Slavnov-Taylor identity 
\begin{equation}
\widetilde{Z}_1(\mu^2,\Lambda^2_\mathrm{UV}) 
= Z_g(\mu^2,\Lambda^2_\mathrm{UV}) \,
\widetilde{Z}_3(\mu^2,\Lambda^2_\mathrm{UV}) \, 
Z_3^{1/2}(\mu^2,\Lambda^2_\mathrm{UV})\,,
\label{STI} 
\end{equation} 
In Landau gauge, the ghost-gluon
vertex is ultraviolet finite, and therefore one always has the choice
of setting $\widetilde{Z}_1 = 1$ \cite{Taylor:1971ff}. Thus for a
fixed $Z_g$, corresponding to a fixed choice of the renormalised
coupling $\alpha(\mu^2)=g^2(\mu^2)/4\pi$, one has a
unique relation between the renormalisation factors $\widetilde Z_3$
and $Z_3$ of the ghost and gluon propagators.  In terms of
renormalisation conditions this corresponds to the relation 
\begin{equation}
G(\mu^2)^2 Z(\mu^2)=1, \label{renorm} 
\end{equation} 
i.e. the ghost and gluon dressing functions may not be renormalised 
independently for a given renormalised coupling \cite{vonSmekal1998}.
Accordingly, the input values $G(s^2)$ and $Z(s^2)$ have to be chosen to
satisfy (\ref{renorm}). In the infrared analysis in
Sec.~\ref{sec:sana} these could be taken from the well-known infinite 
volume infrared solutions in accordance with the renormalisation 
conditions \eq{eq:relRG}. For a numerical solution, however, this is
not an option.

The simplest way to implement (\ref{renorm}) would be  
to choose $s^2=\mu^2$, fix $Z(s^2)$ and $G(s^2)=1/Z(s^2)^{1/2}$.
Unfortunately this is not possible: the function $g \rightarrow \mu^2$
is not known before the equations are solved completely and
accordingly choosing a specific $g$ is not sufficient to determine the
value of $\mu^2$. One thus chooses an arbitrary, usually large,
subtraction point $s^2$, fixes $Z(s^2)$ and then solves for a range of
values for $G(s^2)$. In the infinite volume limit one
obtains continuous and differentiable solutions only for an extremely
narrow range of values $G(s^2)$ around the correct one implementing
(\ref{renorm}) \cite{fa}.

However, this is no longer true on the compact manifold. We found that
for given values of $g(\mu^2)$ and $Z(s^2)$ one can generate a
continuous array of solutions by varying $G(s^2)$.  These solutions
all behave different in the infinite volume limit. In previous works
\cite{Fischer,Fischer:2005ui} these ambiguities have been resolved by
reading off $Z(s^2)$ and $G(s^2)$ at a large $s^2$ from the infinite
volume solution. However, this procedure is not sufficient. Since one
works with a fixed cutoff $\Lambda_\mathrm{UV}$ on the torus, one
encounters $O(1/\Lambda_\mathrm{UV})$ effects in the ultraviolet
momentum region, similar to $O(a)$-effects in lattice QCD. Thus the
ultraviolet behaviour of the torus solutions is slightly different
from the infinite volume results. This difference is significant for the
renormalisation procedure and in turn affects the infrared behaviour
of the solutions and also the scaling behaviour with volume.

To avoid these UV-cutoff effects we modify the RG procedure by
utilising the analytic results in Sec.~\ref{sec:sana}: the RG
condition on the torus is adjusted to approach the infinite volume RG
condition in the limit of very large volumes as implemented in
Sec.~\ref{sec:sana} in (\ref{eq:gluonren}) and (\ref{eq:ghostren}). To
that end we choose $G(s^2)$ for fixed $g$ and $Z(s^2)$ accordingly. As
a result we find that the values of $Z(s^2)$ and $G(s^2)$ are
different but still close to the corresponding values of the infinite volume 
solution.

To summarise: we find that $O(a)$-effects have to be accounted for in
the renormalisation procedure for DSEs on the torus to obtain the
correct infinite volume limit. This technical improvement compared to
previous works \cite{Fischer,Fischer:2005ui} allows us in turn to
reliably explore the large volume behaviour of the ghost and gluon
propagators.

\subsection{Propagators \label{sec:props}}

\begin{figure}[th!]
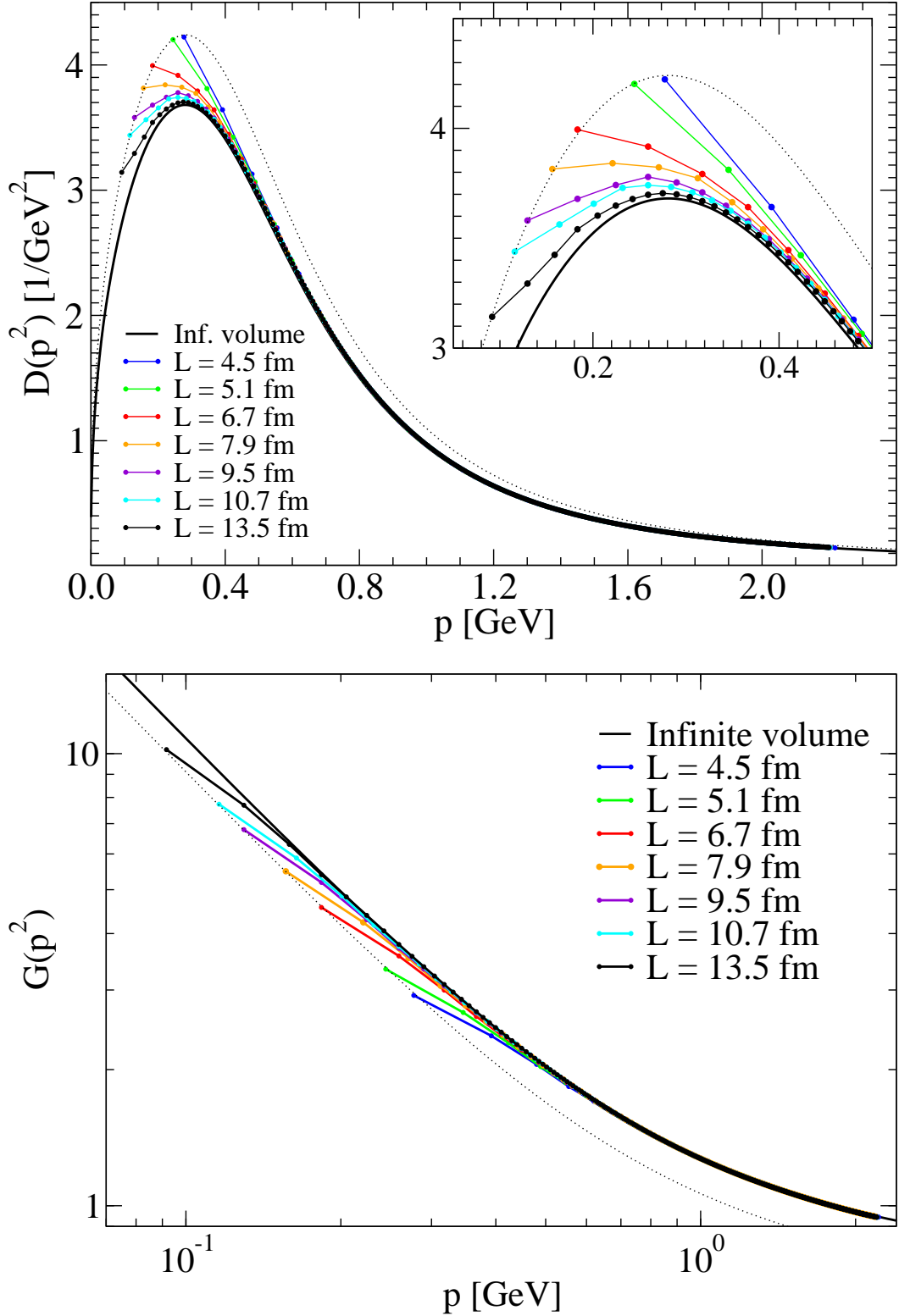

\centerline{
\epsfig{file=Fig3a.eps,width=14cm}}
\vspace*{4mm}
\centerline{
\epsfig{file=Fig3b.eps,width=14cm}}
\caption{Numerical solutions on tori with different volumes compared
  to the infinite volume limit. The upper graph shows the
  gluon propagator, whereas on the lower graph the ghost dressing
  function is depicted.}\label{Fig2}
\end{figure}

Our numerical results for the ghost and gluon propagator on different
volumes are shown in the two graphs of Figure \ref{Fig2}. The momentum
scale is fixed by comparison with corresponding lattice calculations:
we demand that the maximum of the gluon dressing function in the
DSE-approach occurs at the same momentum scale as in the lattice
dressing function. Hereby we assume implicitly that truncation errors
due to the neglect of the gluonic two-loop diagrams only mildly affect
the position of the maximum. As a crosscheck we compare the resulting
values of the DSE and the lattice gluon propagator at the lowest
momentum point accessible on manifolds of similar volume. The result,
shown explicitly in Sec.~\ref{sec:lattice} below, suggests that
scale uncertainties relative to the lattice scale might roughly be of
the order of 10 percent.

We discuss results on seven different volumes $V=L^4$; the
corresponding box lengths $L$ are given in the legends of Figure
\ref{Fig2}. One clearly observes that the infinite volume solutions of
the gluon propagator $D(p^2)$ and ghost dressing function $G(p^2)$ are
more and more approached by the torus solutions with increasing
volume. This tendency will be quantified in Sec.~\ref{sec:asym}.
Qualitatively one can see the following behaviour from the plot: the
propagator seems to be divergent at volumes of $V \approx (4-8
\,\mbox{fm})^4$. Note that these volumes are already large compared to
the ones used in most lattice calculations of observables. For even
larger volumes the propagator bends downwards to reach a plateau at
roughly $V \approx (9 \,\mbox{fm})^4$.  For volumes larger than $V
\approx (10 \,\mbox{fm})^4$ the propagator is infrared vanishing and
therefore qualitatively similar to the infinite volume limit given in
(\ref{powers}) and (\ref{kappa}).

This finding becomes even more pronounced when one applies momentum
cuts: from the inset of the graph for the gluon propagator one can see
that the first three momentum points on the dual torus behave
differently than the other points. In the language of
Sec.~\ref{sec:mom} this means that the 'angular sum' over $j$ on these
spheres is by no means a good representation of the corresponding
angular integrals in the infinite volume limit. If one omits these
points in the representation of $D(p^2)=Z(p^2)/p^2$ and $G(p^2)$, then
the turnover of the gluon propagator shifts from roughly $V \approx (9
\,\mbox{fm})^4$ to $V \approx (11 \,\mbox{fm})^4$. In such a
representation only the propagator of our largest volume, $V = (13.8
\,\mbox{fm})^4$, can be seen as infrared vanishing. Taken at face
value this means that contemporary lattice calculations are still far
away from the critical volume where the behaviour of the infinite
volume solution can be deduced.

Similar conclusions can be drawn from the results for the ghost
propagator. In Figure \ref{Fig2} we observe that the first two or
three points on each result for the ghost bend away from the power law
behaviour of the infinite volume solution.  The remaining curves have
more and more points on the 'scaling region' of the ghost dressing
function, where the infinite volume power law develops. The critical
volume, where the full infinite volume power law can be seen on the
torus, is of roughly the same size as the one for the gluon
propagator.

\begin{figure}[th!]
\centerline{
\epsfig{file=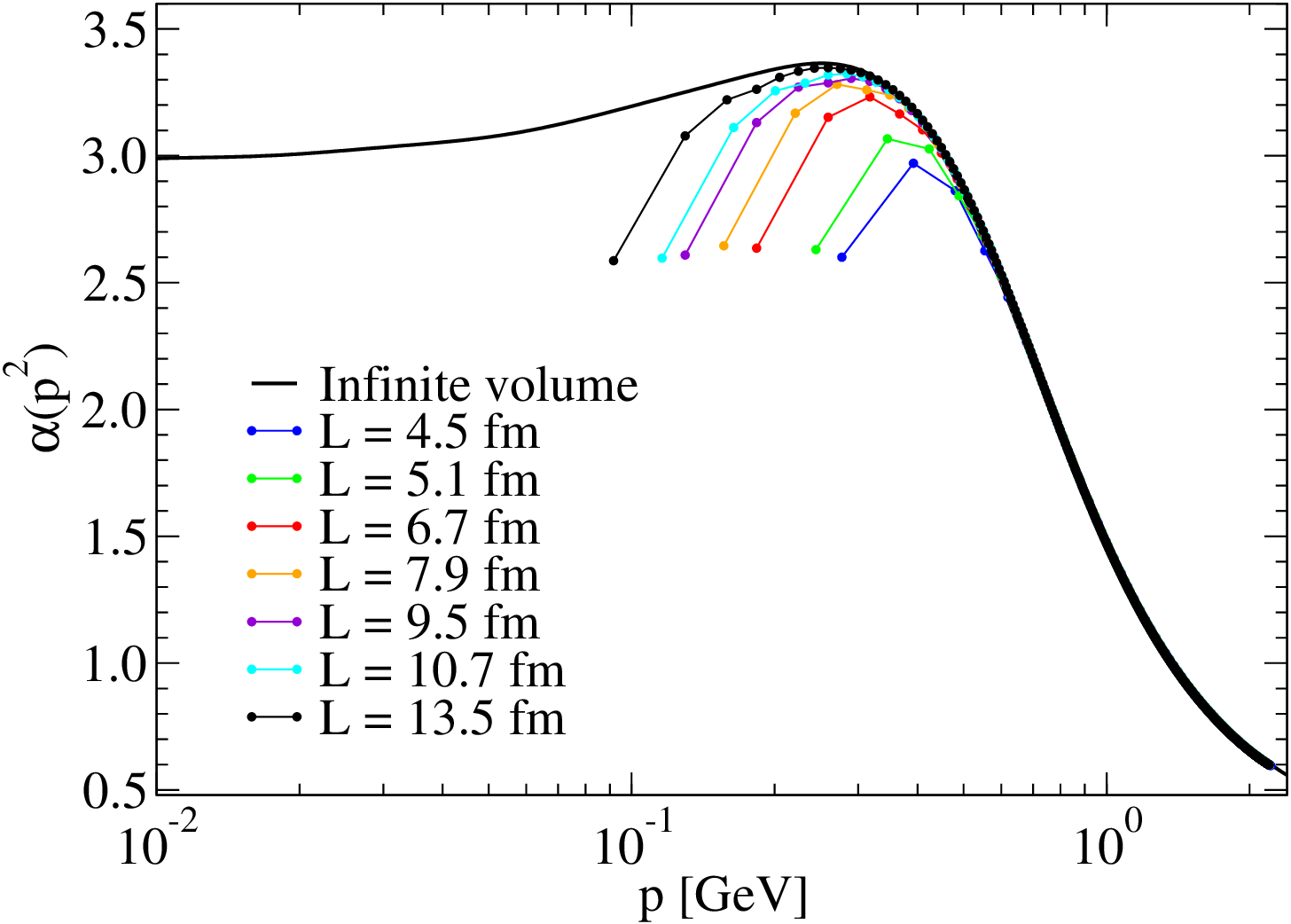,width=13cm}}
\caption{The running coupling $\alpha(p^2)=\alpha(\mu^2) G^2(p^2)
  Z(p^2)$ as defined from the ghost-gluon vertex. \label{Fig3}}
\vspace*{8mm} \centerline{ \epsfig{file=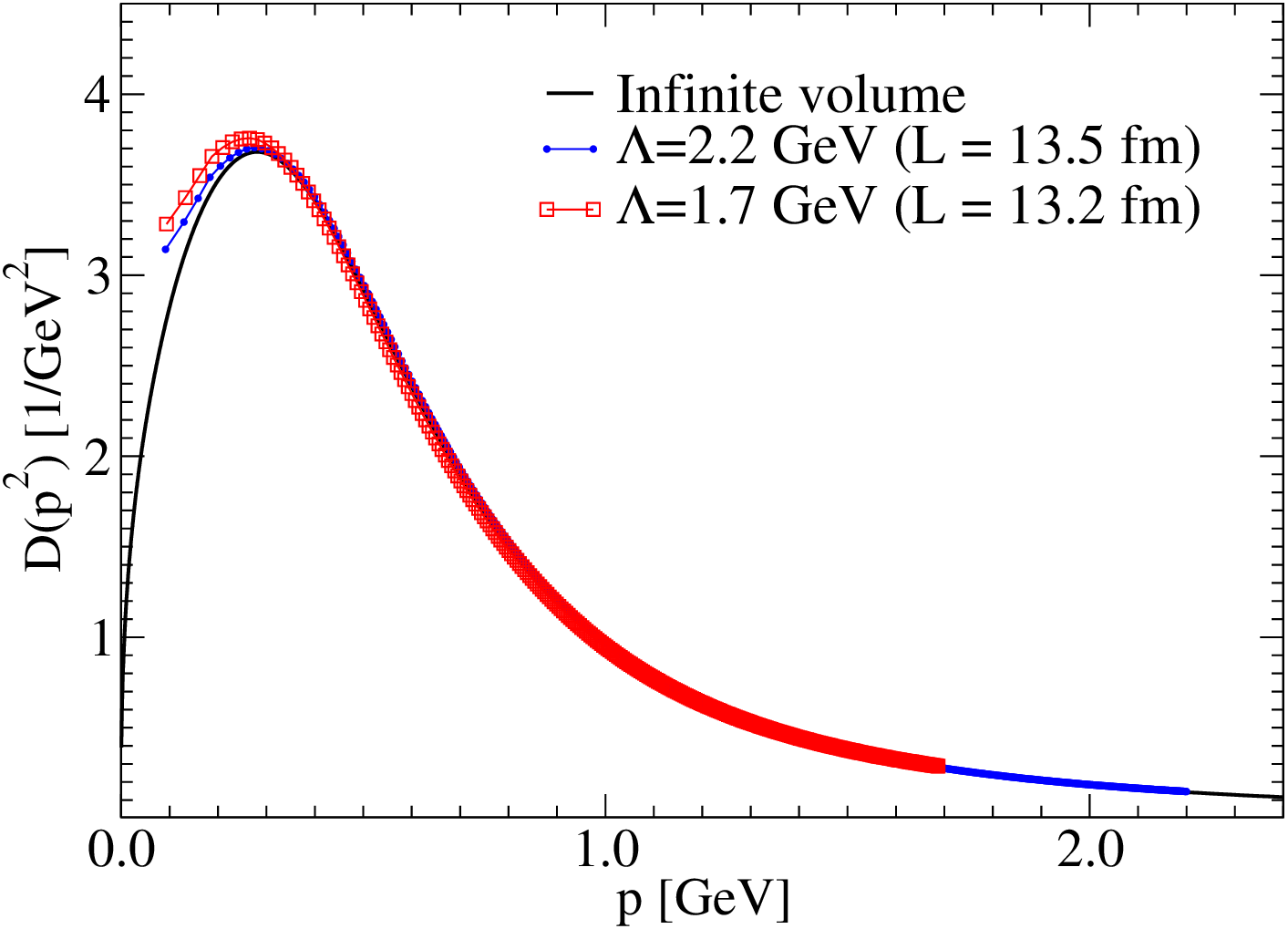,width=13cm}}
\caption{The gluon propagator calculated on tori with the same
  physical volume but different ultraviolet cut-offs. \label{Fig3a}}
\end{figure}

It is interesting to note that the lowest momentum point of the gluon
and ghost solutions follow a curve distinct from the infinite
volume solution. It turns out that in both cases this curve
can be described by the infinite volume solution multiplied
by a constant factor (the dotted lines in the graphs). A similar
observation can be made for the second lowest point. As yet we have no
full analytical understanding of this observation.

This behaviour has interesting consequences for the running coupling,
defined in eq.~(\ref{coupling}) and shown in Figure \ref{Fig3}: at
least the first two points of every solution on a torus will always
remain significantly below the infinite volume solution no matter how
large the volume is. Thus on every compact manifold one finds a
running coupling which looks infrared suppressed, although the
corresponding coupling from the infinite volume limit has an infrared
fixed point. Taken at face value this result means that it is
extremely difficult if not impossible to verify infrared fixed points
of the running coupling in lattice calculations. Indeed, all recent
determinations of vertex couplings from lattice QCD find infrared
vanishing \cite{Boucaud:1998xi} or strongly suppressed couplings 
\cite{Furui:2003jr} whereas the infinite volume analysis unambiguously 
predicts fixed points \cite{vonSmekal1997,Alkofer:2000wg,afl,Fischer:2006vf}.  
Our observation may well serve to explain this discrepancy.

Finally, in the lower panel of Figure \ref{Fig3a} we present results
for two tori with the same physical volume but different ultraviolet
cut-offs given in the legend. The two corresponding solutions for the
gluon propagator are close to each other, although one observes a
slight distortion of shape from the curve with the small cut-off
compared to the one with the larger cut-off and the infinite volume
solution.  From this result one may conclude that cut-off effects in
the DSE-solutions on a torus are small and do not significantly affect
the infrared behaviour of the propagators.\footnote{In the light of
  these results we also conclude that the larger cut-off effects noted
  in \cite{Fischer:2005ui} are artefacts of the renormalisation
  procedure used therein.}

\subsection{Emergence of the infrared asymptotic behaviour
  \label{sec:asym}}

We now study the rate of approach towards the infinite
volume solution in some more detail.  Because of the
redundancy in the parametrisations (\ref{eq:expand}) as discussed in
Sec.~\ref{sec:ssvols}, direct fits to our data simultaneously of all free
parameters in the finite-volume infrared forms are not possible.
Fits of both screening-mass parameters would in principle be possible
from the expressions in Eqs.~(\ref{eq:IRfit}). It will turn out,
however, that the volumes in our numerical results which range up to
approximately (14 fm)$^4$ are nevertheless still too small for the
leading infrared exponent $\kappa_c$, or any other momentum and volume
independent constant, to describe the data in a reasonable range of
momenta. For these volumes the available infrared momenta all lead to
values of $1/(k^2L^2)$ which are beyond the range of validity of the
expressions in (\ref{eq:IRfit}), where the exponent $\kappa_c$ is
already close to its infinite volume value but the screening masses
$\mu_{Z/G}$ are still present. We will find, in fact, that there is
not yet a clear separation of scales (\ref{eq:valrange}) for any
momentum in these volumes in the first place.

\begin{figure}[t]
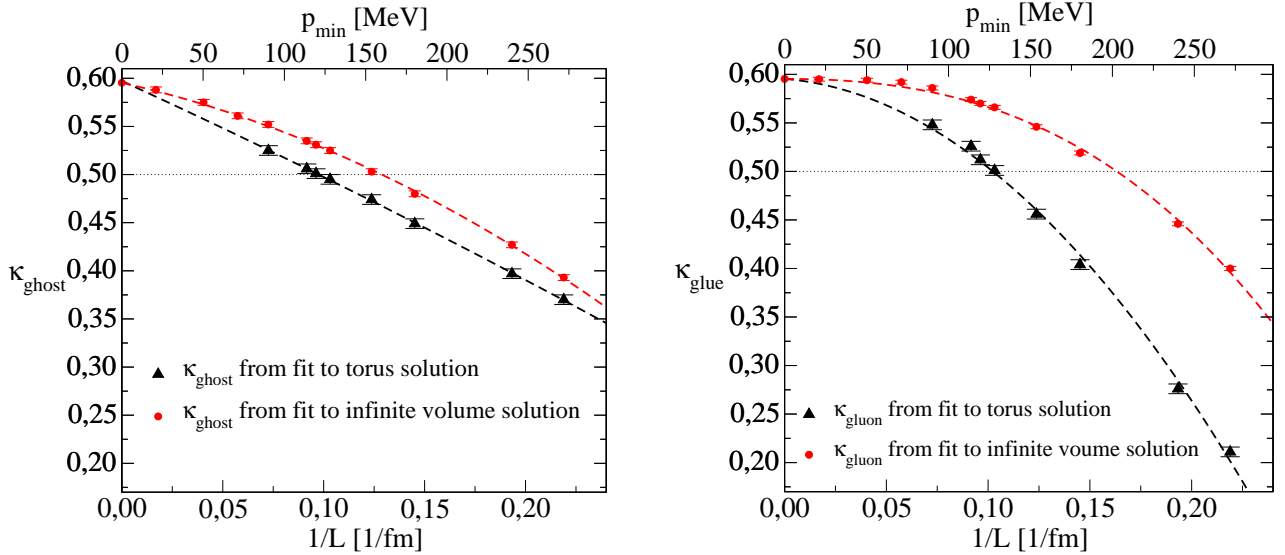

\centerline{
\epsfig{file=Fig6a.eps,width=8cm} \hfill \epsfig{file=Fig6b.eps,width=8cm}}
\caption{The infrared exponents $\kappa_{ghost}$ and $\kappa_{glue}$
  from fits to the torus solutions on different volumes, {\it c.f.},
  model (1) as described in the text, compared to the effective power
  law fits to the infinite volume solution in the same momentum region
  for model (2).
  \label{fig:fit}}
\end{figure}

\begin{figure}[t]
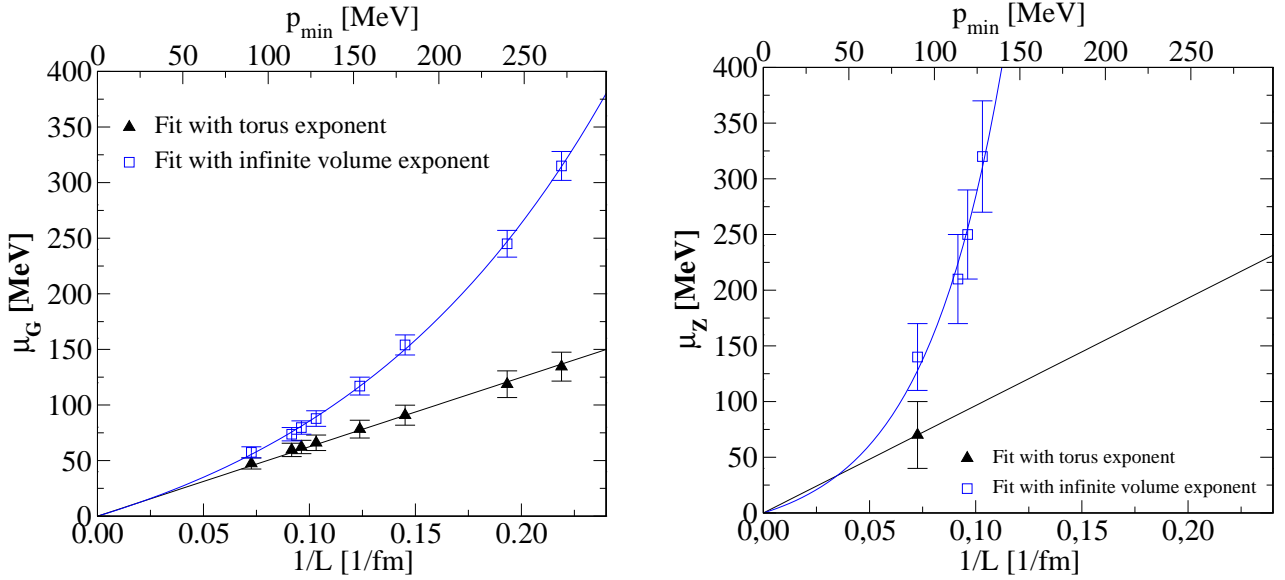

\centerline{
\epsfig{file=Fig7a.eps,width=8cm} \hfill \epsfig{file=Fig7b.eps,width=8cm}}
\caption{In the left panel the behaviour of the ghost mass $\mu_G$
  with the inverse box length $1/L$ is shown.  The error of five
  percent reflects the systematic uncertainty of our fit procedure.
  The straight line is a linear fit with zero offset. In the right
  panel the same is shown for the gluon mass $\mu_Z$.  However, the
  systematic uncertainty is larger, about forty percent, see text.
  \label{fig:4c}}
\end{figure}

We therefore adopt two different models, each to fit the data in two
steps.  To illustrate this procedure we start from the ghost dressing
function for which it yields the more stable results in the available
volumes than for that of the gluon. Instead of (\ref{eq:IRfit}) we
assume a ghost dressing function at low momenta of the form,
\begin{eqnarray}
  G_L(k^2)& = & G_\mathrm{IR}
  \left(\0{\Lambda^2_\mathrm{QCD}}{ k^2+\mu_G^2}
  \right)^{\kappa_{ghost}} \,, 
\label{ghostfit}
\end{eqnarray}
in which we allow both, the screening mass $\mu_G$ and the effective
exponent $\kappa_{ghost}$ to depend on the volume, {\it i.e.}, on $L$.
We then proceed in the following two alternative ways:

\begin{enumerate}
\item In this model we first set the screening mass to zero, {\it
    i.e.}, $\mu_G =0$ in (\ref{ghostfit}), and determine the effective
  exponent $\kappa_{ghost}$ by fitting the third to fifth lowest
  momentum values of the solution for every given volume to the
  resulting pure power law $\propto
  (\Lambda^2_\mathrm{QCD}/k^2)^{\kappa_{ghost}}$. The results are
  plotted over the inverse box length as the triangles in the left
  panel of Fig.~\ref{fig:fit}.

  \smallskip\noindent We then keep $\kappa_{ghost}$ fixed and fit in a
  second step the remaining parameters on a larger region. Here we
  typically use the lowest eight momenta. The resulting values for
  ghost screening mass are the triangles in the left panel of
  Fig.~\ref{fig:4c} which all lie on a straight line $\propto 1/L$
  with offset zero.
 
\end{enumerate}
  
The resulting fits are stable with respect to moderate variations of
the fit regions in each step (within error bars) and nicely reproduce
the same results.  Note that this method predominantly absorbs the
finite volume effects in the effective exponent. The screening mass
essentially accounts only for the remaining mismatch after this first
step.

\begin{enumerate}
\setcounter{enumi}{1}
\item In this fit-model we adopt the extremely opposite point of view.
  We first obtain an effective exponent $\kappa_{ghost}$ which is
  completely independent of any finite-volume effects by fitting the
  pure power law form to the infinite volume solution in the
  same momentum range used in the first step of model (1) above, that
  of the third to the fifth lowest momentum values of a given torus
  solution.  The resulting values are plotted for comparison as the
  dots which lie above the triangles, also in the left panel of
  Fig.~\ref{fig:fit}.
 
  \smallskip\noindent Fitting the remaining parameters to that torus
  solution then essentially works as in model (1) above. We use the
  same larger momentum region of the typically eight lowest values,
  and the fits are similarly stable w.r.t. variations of that region
  as above.  The resulting screening masses are larger than in model
  (1) and shown as the squares in Fig.~\ref{fig:4c}.
  
\end{enumerate}

The crucial distinction is that in this second model the finite-volume
effects are entirely accounted for by the screening mass alone.  For
the volumes available the effective exponent $\kappa_{ghost}$ in
(\ref{ghostfit}) is not equal to $\kappa_c$ yet. It changes with the
volume because the momentum range where it is obtained does.
Nevertheless, this effective exponent is obtained from the infinite
volume solution and thus independent of finite volume effects. It
rather describes the deviations from the leading infrared behaviour
already of the infinite volume solution in the corresponding momentum range,
due to the fact that these momenta are not yet far enough below
$\Lambda_\mathrm{QCD} $. The volumes are too small to fit a range of
momenta (\ref{eq:valrange}) in.

The fact that both alternative fit-models are possible reflects the
redundancy in our finite-volume infrared parametrisations as mentioned
above. In volumes which are not quite large enough for the leading
volume asymptotics we can neither describe the data with constant
exponent nor without screening mass. Both, an effective exponent and a
screening mass are necessary in these volumes. However, to allow this
both underconstrains the fits. The two extreme models to fit the data
by absorbing a maximum of finite-volume effects either in the
effective exponent or the screening mass serve as a measure of this
redundancy.

The results approach each other and this redundancy decreases with
increasing volumes as expected. In the volumes of the order of up to
approximately (14 fm)$^4$ presented here the mismatch between both
models does not quite vanish yet. We can extrapolate the data from
both fit-models to larger volumes, however, to make a rough prediction
about what volumes are necessary for the unique leading finite-volume
infrared behaviour as parametrised in (\ref{eq:IRfit}) to set in. The
result gives the following quite consistent overall picture:
   
Comparing Figs.~\ref{fig:fit} and \ref{fig:4c} we observe that for box
lengths of approximately $L = 40$ fm ({\it i.e.}, $1/L \approx 0.025\;
\mbox{fm}^{-1}$) the effective exponents from torus and infinite
volume solutions approach each other and eventually meet extremely
close to the asymptotic infinite-volume value $\kappa_c \approx 0.595
$. For lengths above 40 fm we therefore expect $\kappa_{ghost} =
\kappa_c$ to be a very good approximation.  The screening masses are
not zero yet, but also approach each other very well at volumes of
this size. This means that from our extrapolations we would estimate
that the asymptotic forms (\ref{eq:IRfit}) can be expected to describe
the finite volume infrared behaviour well for volumes of $L =$ 40~fm
in size and more.

As described in the previous section, the dressing function of the
gluon propagator is more difficult to describe by such fits in the
intermediate volumes here because its momentum dependence deviates
much more strongly from its infinite-volume infrared asymptotics even
at the lowest momenta available in our volumes.  For the gluon
dressing function model (1) only leads to stable results for the
screening mass $\mu_Z$ in our largest volume $V = (13.8 \,{\rm fm})^4$
which is included in the right panel of Fig.~\ref{fig:4c}.  The volume
dependence of the gluon dressing function is stronger than that of the
ghost. This is apparent from the right panel in Fig.~\ref{fig:fit} for
model (1), with the volume dependence predominantly accounted for by
the effective exponent which is obtained from the $\mu_Z = 0$ fits of
the form,
\begin{equation}
  Z_{L}(k^2) =
  Z_{\rm IR} \, \left(\frac{k^2}{\Lambda_{\rm QCD}^2}\right)^{2
    \, \kappa_{glue}} \; , \label{gluonfit}  
\end{equation} 
and which depends more strongly on $1/L$ than it does for the ghost.
But it again approaches the effective exponent of the infinite volume
solution at box lengths of about 40 fm.

In model (2), with no finite-volume effects in the effective
exponents, we observe the same stronger volume dependence in the
gluonic screening mass $\mu_Z$ as compared to $\mu_G$ in
Fig.~\ref{fig:4c}. Again, however, within the considerable
errors,\footnote{For the gluonic screening-mass in model (1) we simply
  draw a straight line between our only value at $L = 13.8 $ fm and
  zero at $1/L = 0$ in Fig.~\ref{fig:4c} (right).} the extrapolations
of the data for the gluonic screening masses from models (1) and (2)
are consistent with an approach of the two at about the same length
scale of $L = 40$ fm.

Note that the screening-masses of model (1) are extrapolated using
linear fits (of the fits) inversely proportional to the length,
\begin{equation} 
  \mu_{Z/G} \, \propto \, 1/L  \; .
\end{equation}
In model (2) on the other hand, our extrapolations of the screening
masses are exponential in nature from fits of the form
\begin{equation} 
  \mu_{Z/G} \, \propto m \, \big(\exp\{l/L\} - 1  \big) \; , 
\end{equation}
with dimensionful constants $m$ (in MeV) and $l$ (in fm).  In our
intermediate volumes both forms differ considerably. At sufficiently
large $L$ (with our estimate $\ge 40$ fm), however, they both agree
with the scaling Ansatz (\ref{muLim}) used the infrared analysis of
Sec.~\ref{sec:ssvols}.

As a final cross-check, for the ratio of the screening masses,
\begin{equation} 
  {\mu_Z}/{\mu_G} \approx \sqrt{238/86} \approx 1.7 \; , 
  \label{scmrat}
\end{equation}
as obtained from Eqs.~(\ref{c-rel}) in Sec.~\ref{sec:ssvols} for
asymptotically large volumes, here we roughly obtain ${\mu_Z}/{\mu_G}
\approx 1.6 $ for their ratio from the corresponding ratio of the
slope of the linear fits in model (1), the solid lines through the
triangles in Fig.~\ref{fig:4c}. This is in perfect agreement with the
predicted asymptotic ratio (\ref{scmrat}), but the procedure has a
large error of around 50\%, mainly from the single value $\mu_Z =
70(30)$ MeV for the $(13.8 \,{\rm fm})^4 $ volume which determines
the slope in the gluonic screening mass in model (1).

For the effective exponents of both dressing functions we always use
extrapolations (the long dashed lines for model (1) and the short dashed
lines for model (2) from the infinite volume solutions) based on quadratic
polynomials\footnote{The only exception here is the effective exponent
  of the infinite-volume gluon solution, the short dashed line in the
  right panel of Fig.~\ref{fig:fit}, which requires the inclusion of a
  cubic term.}  in $1/L$ with the $1/L = 0 $ value fixed at
$\kappa_c\approx 0.595$,
\begin{equation}\label{ffit}
  \kappa_{fit}(1/L) 
  = \kappa_c  - \frac{a_1}{L} - \frac{a_2}{L^2}  \; . 
\end{equation} 

For the dressing functions in model (1) we have also included the
constant term $\kappa_{fit}(0) $ in the quadratic fit of the form
(\ref{ffit}):
 From the ghost dressing function this yields $\kappa_{fit}(0) =
0.597(3)$ in perfect agreement with the analytic result of $\kappa_c =
0.595$ at 3-digit precision.

Note that for the gluon dressing function, the model (1) fits using
the form (\ref{ffit}) with fixed $\kappa_c$ produce a coefficient
$a_1$ of the linear term consistent with zero. Including the constant
term in the complete 3 parameter fit is then not stable anymore. It
requires dropping the linear term, {\it i.e.} freezing $a_1=0$. Doing
so, however, one then obtains $\kappa_{fit}(0) = 0.60(2)$ which is
also consistent with the analytic result.

Finite volume corrections to $\kappa_c$ of the form (\ref{kappaLim})
used for our infrared analysis in Sec.~\ref{sec:ssvols} can not reasonably
be extracted from our present data. This is not unexpected because it
requires volumes of a size at which the scaling relation
(\ref{kapparel}), {\it i.e.}, $\kappa_{Z_L} + 2\kappa_{G_L} = 0$, for
the infrared exponents is already valid. Here, {\it c.f.}, the
definition of the effective exponent for our fits in (\ref{gluonfit}),
this is when
\begin{equation}
  \kappa_{glue} \approx \kappa_{ghost} \; ,
\end{equation} 
which according to our estimate again requires $L \approx 40$ fm, {\it
  c.f.}, Fig.~\ref{fig:fit}, and is thus still relatively far beyond
our numerical data to attempt reasonable fits and extrapolations.

We furthermore note that the length scale where the behaviour of the
gluon propagator turns from $\kappa<0.5$ (infrared diverging) to
$\kappa>0.5$ (infrared vanishing) is of the order of at least $L= 6$
fm for model (2) (from the infinite volume effective exponent) up to
$L = 10 \,\mbox{fm}$ for model (1). The exact turning point might also
be somewhat truncation dependent.

In summary, we have seen in the previous two sections that the
qualitative infrared behaviour of the gluon and ghost propagators of
Landau gauge QCD starts to emerge in volumes of estimate sizes between
$L = 10$ fm and 15 fm. To reliably extract infrared exponents and
other quantitative results about their infrared behaviour, one
furthermore needs a certain range of momenta with a clear separation
of scales (\ref{eq:valrange}) which requires much larger volumes.  The
leading finite-size effects are under good control, and can be
described by the simple screening masses $\mu_{Z/G} \propto 1/L$ via
Eqs.~(\ref{eq:IRfit}), when the effective exponents approach their
infinite volume scaling behaviour and are sufficiently close to
$\kappa_c$. Our extrapolation method predicts that this might require
volumes of about 40 fm in size.

\subsection{Comparison to lattice results \label{sec:lattice}}

\begin{figure}[t]
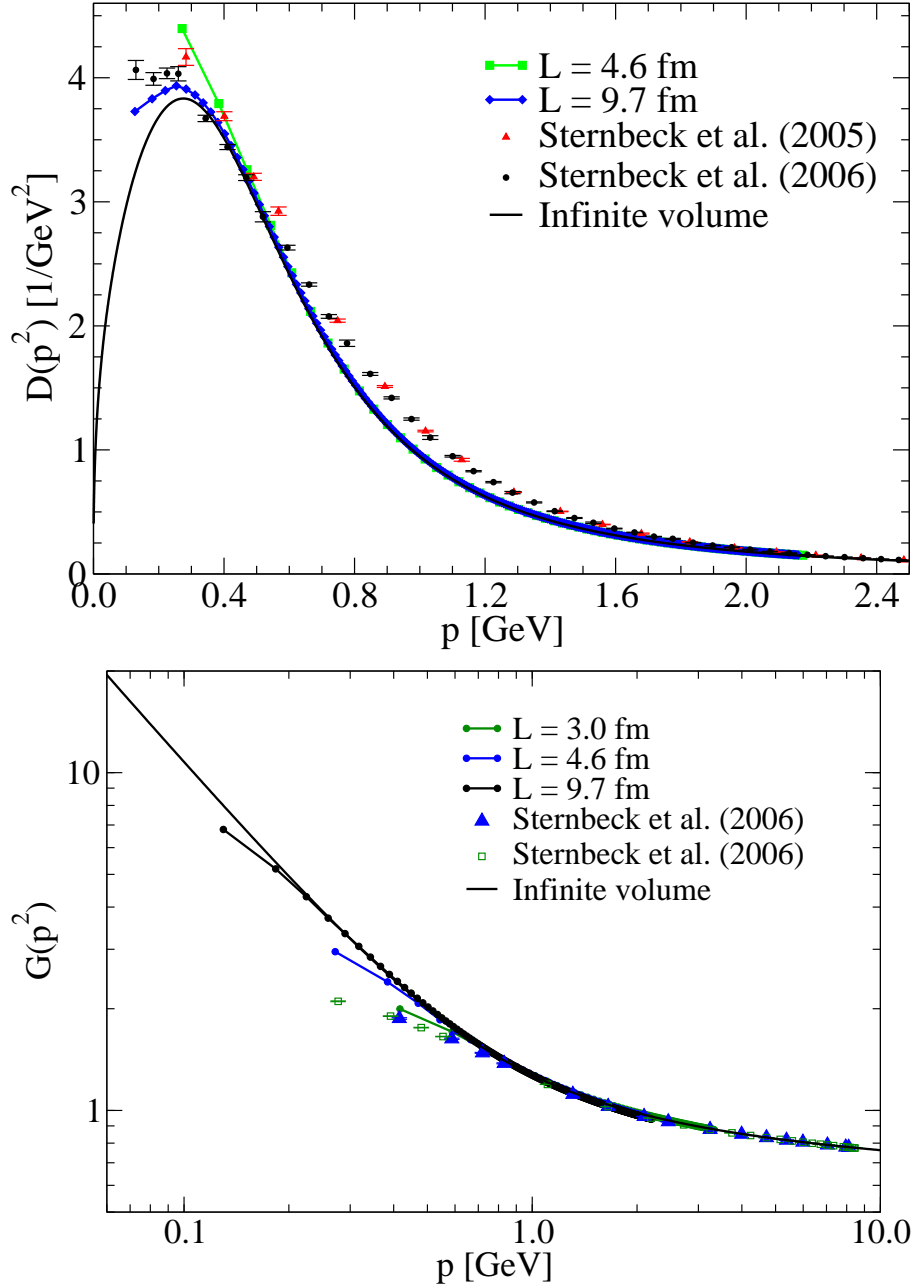

  \centerline{ \epsfig{file=Fig8a.eps,width=12cm}} \vspace*{2mm}
  \centerline{ \epsfig{file=Fig8b.eps,width=12cm}}
\caption{DSE results for the gluon propagator and ghost dressing
  function on tori with different volumes compared to recent lattice
  calculations on similar manifolds. The lattice data are taken from
  Refs.~\cite{Sternbeck:2006cg,Sternbeck:2005tk}.
\label{fig:latt}}    
\end{figure}

We now compare our results from DSEs on a torus to the ones from
lattice calculations. Lattice studies of the Landau gauge gluon
propagator have quite a long history already
\cite{Man87,Ber93,Mar95,Ais96,Lei98}. Nowadays, lattice data for the
gluon propagator is available on impressively large lattices. The
authors of \cite{Cucchieri:2006xi} report on an $SU(2)$-study on a
$52^4$-lattice, whereas in \cite{Sternbeck:2006cg} results from an
$SU(3)$ calculation on a $56^4$-lattice are discussed with larger
lattices on their way.\footnote{For results from simulations using
  large asymmetric lattices, see \cite{Sil04}.} The physical volumes
of both these studies are comparable. The resulting propagators are
very similar with small deviations for the first few points in the
infrared.\footnote{This independence of the number of colours has also
  been found in an earlier study \cite{Alexandrou:2002gs}. However,
  there a different gauge, the LaPlacian gauge, has been used.}  In
the upper graph of Figure~\ref{fig:latt} we display the
$SU(3)$-results together with data on a smaller volume
\cite{Sternbeck:2005tk} and compare with DSE-results on tori with
similar volumes.

The qualitative agreement of the solutions at similar volumes in the
infrared is striking.\footnote{Differences in the intermediate
  momentum regime around 1 GeV are truncation artefacts of the DSE
  solutions. One can show analytically that this is exactly the region
  where the omitted gluonic two-loop diagrams contribute
  significantly.}  Whereas both, the lattice and the DSE result at the
smaller volume $V \approx (4.6 \,\mbox{fm})^4$ seem to diverge, one
starts to observe an infrared finite one, or perhaps even a slight
infrared suppression, at the larger volumes $V \approx (9.7
\,\mbox{fm})^4$.  This indicates that the scaling behaviour of the
lattice results with volume may be very similar to the ones of the DSE
solution.  If so, then our results from the last section predict
also a turnover of the lattice data when even larger volumes are
considered.

The situation is less clear for the ghost dressing function, which was
first studied on the lattice in \cite{Sum96}. Our results for three
different volumes, $V = (3.9, 4.6, 9.7 \,{\rm fm})^4$, are compared to
the SU(3) lattice results of
\cite{Sternbeck:2006cg,Sternbeck:2005tk}.\footnote{For corresponding
  results in $SU(2)$ see \cite{Gattnar:2004bf}.  To date no systematic
  lattice study on the effects of the gauge group on the ghost
  propagator exists.}  For the DSE solutions we observe a
characteristic deviation of the two lowest momentum points at each
volume from the infinite volume solution. As shown in the previous
section these deviations correspond to a ghost mass which goes to zero
in the infinite volume limit. The lattice results do not seem to show
such behaviour as yet. Even though the lattice volumes herein are
roughly between 3 fm and 4.5 fm, and thus still rather small compared
to our analysis in the previous section, there appears to be not
much sign of a volume dependence at all at this point.  The situation
is indeed reminiscent of the corresponding torus solution from the
previous DSE studies. Perhaps revisiting the renormalisation procedure
as described in Sec.~\ref{sec:renorm} might have a similar influence
in the analysis of the lattice data.

In these relatively small volumes, however, some finite-volume effects
in our DSE solutions might also still be somewhat truncation dependent
and could therefore contribute to the differences between the DSE
solutions and on the lattice simulations. Furthermore, effects from
Gribov copies are known to influence the lattice results for the ghost
propagator much stronger than for the glue. This has been investigated
by comparing results from gauge fixing to the first Gribov copy ({\it
  fc}) to the results of a more involved procedure of repeated gauge
fixing and selecting then the best copy ({\it bc}). On rather small
volumes these two procedures have been compared in
\cite{Sternbeck:2005tk} and noticeable effects observed. Similar conclusions 
have been drawn in \cite{Oliveira:2006yw}. Therefore, a
meaningful comparison between the DSE and lattice ghost propagator not
only awaits an improved truncation scheme in the DSEs and results on
larger volumes on the lattice, but also a further clarification of
gauge fixing effects in both, the continuum and lattice studies.

\section{Conclusions}\label{sec:sconc}

The results presented here explicitly demonstrate that Landau gauge
propagators from lattice and continuum calculations are compatible,
provided finite volume effects are properly taken into account. The
approach of the torus solutions towards the infinite volume limit is
slow, but steady and quantifiable.  Our results show that the
characteristic maximum in the gluon propagator should be measurable in
volumes of the order $V \approx (10-15 \, {\rm fm})^4$ which come into
grasp for lattice calculations. This important qualitative effect will
be the hallmark of the onset of the asymptotic infrared region.

The results also demonstrate, however, that extremely large volumes
will be necessary for a more quantitative comparison between results
in the continuum and on the lattice. The reason is, of course, the
requirement \pref{eq:valrange} to observe the asymptotic infrared
behaviour. Hence, even verifying the scaling of the infrared exponent
according to figure \ref{fig:fit} requires spanning an enormous range
of volumes. Thus the agreement between both methods can be at best
qualitative for quite some time to come. That so far not even a very
good qualitative agreement could be observed can also be understood
with the results presented here. The most characteristic qualitative
feature of the gluon propagator is its maximum. To observe it, it is
trivially necessary to also observe its infrared suppression. This
suppression depends on the quantitative value of the infrared exponent
$\kappa_{glue}$. In particular, its value in Landau gauge, compared,
{\it e.\ g.}, to Coulomb gauge, leads to a very weak infrared
suppression.  Hence very large volumes are necessary to observe this
suppression, and hence the maximum.

Nonetheless, if a clear maximum emerges in the gluon propagator on
larger lattices, confirming the predictions made here, it will
definitely be an important piece in our understanding of the
confinement mechanism in Landau gauge.

Finally we would like to remark that all volume effects discussed here
are only typical for correlation functions involving intrinsically
massless fields. Objects, which have an intrinsic mass-scale, like
quarks or hadrons, will only indirectly be affected by such finite
volume effects. For quarks these effects may be still significant, as
can be seen from the results of \cite{Fischer:2005nf}. The situation
is different, however, for hadrons: the specific far infrared
behaviour of the gluon correlation functions are expected to be nearly
irrelevant to hadronic observables
\cite{Alkofer:2000wg,Fischer:rev,Holl:2006ni}.

\smallskip
{\bf Acknowledgements}\\
We are grateful to Reinhard Alkofer for useful discussions. This work
has been supported by the Deutsche Forschungsgemeinschaft (DFG) under
contracts Fi 970/7-1, GI 328/1-2, and MA 3935/1-1.



\end{document}